\documentclass[useAMS,usenatbib]{mn2e}

\usepackage{graphicx}

\title[
Modeling the effects of dust evolution on the SEDs of
galaxies]{Modeling the effects of dust evolution on the SEDs of
galaxies of different morphological type}
\author[A.P.Schurer et al.]{A Schurer$^{1,2}$\thanks{E-mail:
schurer@sissa.it (APS)}, F. Calura $^3$, L. Silva $^3$, A. Pipino
$^4$, G.L. Granato $^3$,
\newauthor F. Matteucci $^{3,5}$ and R. Maiolino $^6$.\\
$^{1}$Astrophysics Sector, SISSA/ISAS, Via Beirut 2-4, I-34014 Trieste, Italy\\
$^{2}$INAF, Osservatorio Astronomico di Padova, Vicolo
dell'Osservatorio 5, I-35122 Padova, Italy\\
$^{3}$INAF, Osservatorio Astronomico di Trieste, Via G. B. Tiepolo 11 I-34131 Trieste, Italy. \\
$^{4}$Astrophysics, University of Oxford, Denys Wilkinson
Building, Keble Road, Oxford OX1 3RH, UK\\
$^{5}$INAF, Dipartimento Astronomico-Universita di Trieste, Via G. B. Tiepolo 11 I-34131 Trieste, Italy. \\
$^{6}$INAF – Osservatorio Astronomico di Roma, via di Frascati 33,
Monte Porzio Catone, Italy}
\begin{document}
\date{}
\pagerange{\pageref{firstpage}--\pageref{lastpage}} \pubyear{2008}
\maketitle
\label{firstpage}
\begin{abstract}
We present photometric evolution models of galaxies, in which, in
addition to the stellar component, the effects of an evolving dusty
interstellar medium have been included with particular care.
Starting from the work of Calura, Pipino \& Matteucci (2008), in
which chemical evolution models have been used to study the
evolution of both the gas and dust components of the interstellar
medium in the solar neighbourhood, elliptical and irregular
galaxies, it has been possible to combine these models with a
spectrophotometric stellar code that includes dust reprocessing
(GRASIL) (Silva et al. 1998) to analyse the evolution of the
spectral energy distributions (SED) of these galaxies. We test our
models against observed SEDs both in the local universe and at high
redshift and use them to predict how the percentage of reprocessed
starlight evolves for each type of galaxy. The importance of
following the dust evolution is investigated by comparing our
results with those obtained by adopting simple assumptions to treat
this component.
\end{abstract}
\section{Introduction}
Small solid grains of material present within the interstellar
medium (ISM) of galaxies, often referred to as dust, absorb and
scatter stellar radiation mainly in the UV and optical re-emitting
it in the IR. This process has been receiving increasing interest
due to the development of sophisticated satellites operating in the
IR (initially \emph{IRAS}, \emph{ISO} and more recently
\emph{Spitzer}) and large earth bound telescopes (for example
\emph{SCUBA}, a detector on the James Clerk Maxwell Telescope
operating in the sub-mm) which have indisputably demonstrated the
importance of these particles. Indeed it has been found that
approximately 30 per cent of all light radiated by stars in the
local universe has been dust reprocessed (Soifer $\&$ Neugebauer
1991) and measurements of the FIR/sub-mm background estimate a
potentially larger percentage $\sim$ 50 per cent reprocessed over
the entire history of the universe (see e.g. Hauser $\& $ Dwek 2001
for a review).

Hence, a proper treatment of the dust reprocessing in galaxies is
essential, both to derive physical quantities such as star-formation
rates and star-formation histories from observations and also when
testing theoretical galaxy formation models against observations. To
do this it is necessary to model the effect that the ISM has on the
propagation of photons from the stellar sources to the edge of the
galaxy. As a result a good knowledge of the properties of the dust
component and its distribution within a galaxy is required.

Observations from our own galaxy have provided insights into the
composition of interstellar dust, its morphology, size distribution
and relative abundance of its various elemental constituents and
many viable models have been proposed to explain them. The most
commonly used dust compositions include those of Mathis, Rumple \&
Nordsieck (1977), Draine \& Lee (1984), Draine \& Li (2001) and
(2007),Weingartner \& Draine (2001), Des\`{e}rt et al. (1990),
Zubko, Dwek \& Arendt (2004). Such compositions try to recreate the
amount of extinction observed to local stars, the emission from
local dust clouds and the depletion of elements.

Several galactic dust models have been proposed to follow the
propagation of photons through the ISM (see reviews by Dopita 2005
for starbursts, Popescu \& Tuffs 2005 for normal galaxies and Madden
2005 for irregular galaxies). Such galactic dust models treat this
component with differing geometries and with varying degrees of
sophistication. Some, which solely involve radiative transfer
calculations, have been used to account for the optical appearance
of galaxies (Xilouris et al. 1997,1998,1999; Bianchi 2007), some
have only used dust emission models to account for the FIR/Sub-mm
SEDs (Dale et al. 2001; Draine \& Li 2007) and some which combine
radiative transfer with a dust model in order to self-consistently
model SEDs from the UV to the sub-mm and radio (Silva et al. 1998
hereafter S98; Bianchi, Davies \& Alton 2000; Efstathiou,
Rowan-Robinson \& Siebenmorgen 2000, Efstathiou \& Rowan-Robinson
2003, Tuffs et al. 2004; Piovan et al. 2006).

Many of the proposed models are based on the dust composition models
calculated for the Milky Way, and for simplicity use this same dust
composition for all morphological types of galaxies and for every
stage of their evolution. However, unsurprisingly observations
indicate that dust properties are likely to vary with age and
morphological type. For example observations of very high redshift
and hence young galaxies, (Maiolino et al. 2004) and dwarf irregular
galaxies (eg. Galliano et al. 2005) have been used to infer very
different sizes and compositions of dust grains in these galaxies
than those observed in the Milky Way. So it is unlikely to be
sufficient to model the effects of dust at all redshifts and in
galaxies with very different properties and star formation histories
using only one grain distribution model. It is also common to
calculate a total mass of dust by assuming a dust-to-gas mass ratio
proportional to the metallicity of the galaxy, with the same
constant of proportionality in all environments (e.g. Granato 2000
et al.; Baugh 2005 et al.). An assumption which is far from certain,
for example it has been found that in dwarf irregular galaxies a
lower than expected dust-to-gas has been observed (Hunt, Bianchi \&
Maiolino 2005; Galliano et al. 2003, 2005). Hence it would be
beneficial when modeling the effects of dust reprocessing in a range
of different environments to be able to relax these assumption and
select appropriate dust properties for each galaxy individually.
Unfortunately observational evidence regarding the exact dust
composition outside our own galaxy is very limited. It is therefore
necessary at the present time in order to model SEDs of galaxies of
different type and at different stages in their evolution to rely on
theoretical models which follow the evolution of dust.

The dust grains are believed to be formed in a wide range of
astrophysical processes, particularly in the stellar outflow of
giant and Wolf-Rayet stars. Recent FIR to sub-mm observations of
quasars have revealed large masses of dust at high redshift
(Bertoldi et al. 2003; Robson et al. 2004; Beelen et al. 2006). The
presence of such quantities of dust when the universe was $\leq$ 1
Gyr old before significant dust production from giant stars was
possible, suggests an important role should be played by the
enrichment due to supernovae (SNe) (Morgan \& Edmunds 2003; Maiolino
et al. 2004, Dwek et al. 2007) and the observation of dust in the
remnants of SNe (Dunne et al., 2003, 2008, Morgan et al. 2003 but
see also Krase et al 2004, Wilson \& Batrla 2005 and Sugerman et al.
2006) have provided further evidence for the formation of dust in
these objects. However this process is complicated by the
destruction of SNe dust grains by the reverse shock propagating
through the SNe ejecta. Once injected into the ISM the grains are
subjected to a variety of interstellar processes including, thermal
sputtering in high velocity shocks, evaporation and shattering by
grain-grain collisions and accretion in dense molecular clouds
(MCs).

The challenge facing a dust evolution model is therefore to combine
all of these processes in one model and use it to recreate several
observation particularly the dust depletion levels observed in the
local universe, examples include Dwek (1998), Calura et al. (2008)
(hereafter CPM08), Zhukovska, Gail \& Trieloff (2007).

This work aims to combine the two different areas by using two well
tested models, the chemical and dust evolution model of CPM08 to
follow the chemical and dust grain evolution of galaxies of
different morphology and then use the GRASIL code (S98) to use the
calculated stellar and dust evolution in order to calculate the SED
expected for the galaxy. The importance of following the dust
evolution is tested by comparing these fiducial SEDs with SEDs
generated adopting two common simplifications: That the dust to gas
mass ratio is proportional to the metallicity, and that the dust
composition in all galaxies is equal to that derived for the Milky
Way. The main goal of this paper will be a quantitative comparison
between the SEDs generated using the fiducial model and that
adopting the simplifications. In such a way it is possible to
estimate what errors could theoretically be expected from adopting
such assumptions.

We test our models of the three morphologies (spirals, irregulars
and ellipticals) against observed SEDs in the local universe (Dale
et al. 2007) and the model elliptical galaxies at early times
against observations of their expected high redshift counterparts,
SCUBA galaxies (Clements et al. 2008) and massive post-starburst
galaxies at high redshift (Wiklind et al. 2008). We also use the
models to predict how the percentage of reprocessed starlight
evolves for each type of galaxy.

In this work it was only possible to follow in detail the evolution
of the chemical composition of the dust and not how the size
distribution and chemical makeup of the individual grains evolve, a
potentially large but for now unavoidable limitation. To test this,
alternative size distributions for the irregular galaxy and for the
young starbursting elliptical galaxy were proposed based on observed
extinction curves and the resultant SEDs are presented.

The plan of our paper is as follows: In section 2 the chemical and
dust evolution model is explained, in section 3 the
spectrophotometric dust model is introduced and the choices of
parameters discussed, in section 4 the integration of the two models
is briefly described, in section 5 the calculated SEDs are presented
and are compared to observations, in addition a quantitative
comparison is made between them and the SEDs generated using the
simple dust assumptions, in section 6 the implications of our work
is discussed. Some concluding remarks are given in section 7.
\section{Chemical evolution models}
The chemical evolution models that this paper is based upon are
presented in the CPM08 paper. These models follow the chemical
evolution of several different elements in both the gas and the dust
components of the ISM, of spiral, elliptical and irregular galaxies.
More detailed descriptions of these models can be found in Matteucci
\& Tornamb\`{e} (1987), Matteucci (1994) and Pipino et al. (2002,
2005) for the elliptical galaxies, Chiappini et al. (1997, 2001) for
the spirals and Bradamante et al. (1998) for irregular galaxies. For
convenience the main details will be summarized here.
\subsection{Gas Evolution}
All the models consider only one gas phase. In all models the
instantaneous recycling approximation is relaxed and the stellar
lifetimes are taken into account. The abundance of several of the
elements in the gaseous phase of the ISM (H, He, Li, C, N, O,
$\alpha$-elements, Fe, Fe peak elements and S-process elements) are
followed from the birth of the galaxy. The following processes are
considered.
\begin{itemize}
\item The subtraction of elements from the gaseous phase of the ISM due to star formation.
\item The production in stellar winds in low and intermediate mass
non-binary stars with masses 0.8 - 8 $M_{\sun}$.
\item The production in type Ia SNe by binary stars in
systems with combined masses 3-16 $M_{\sun}$.
\item The production in type II SNe of massive stars of masses 8 - 100 $M_{\sun}$.
\item The increase in primordial elements due to infalling gas.
\item The loss of elements from gas outflows due to galactic wind as
the thermal energy of the gas exceeds its binding energy (only
present in the irregular and elliptical models).
\end{itemize}
Low and intermediate mass stars contribute to the ISM metal
enrichment through quiescent mass loss and a planetary nebula phase
by adopting prescriptions by van den Hoek \& Groenwegen (1997). For
massive stars and type Ia SNe the empirical yields suggested by
Fran\c{c}ois et al. (2004) are adopted.

The star-formation efficiencies and the rate of infall in addition
to the Initial Mass Function (IMF) are selected for each of the
galaxy morphologies in order to reproduce observed chemical
abundances.
\subsubsection{Spirals}
\begin{figure}
%%  % Requires \usepackage{graphicx}
 \includegraphics[type=eps,ext=.eps,read=.eps, width=6.5cm, angle=90]{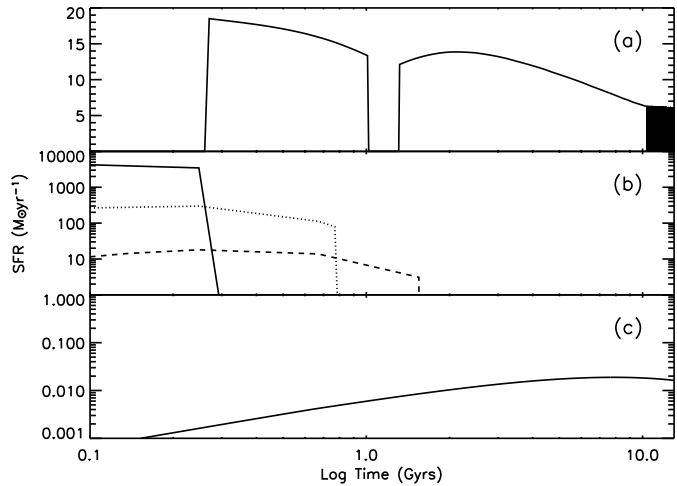}\\
\caption{Predicted SF history in; a) the solar neighborhood ; b)
elliptical galaxies of masses $10^{10} M_{\sun}$ - dashed line,
$10^{11} M_{\sun}$ - dotted line,$10^{12} M_{\sun}$ - solid line; c)
irregular galaxies} \label{a}
\end{figure}

The galactic disc is approximated by several independent rings, 2
Kpc wide, without interchange of matter between them. In this
picture, spiral galaxies are assumed to form as a result of two main
infall episodes. During the first, the halo and the thick disc are
formed. During the second, a slower infall of external gas forms the
thin disc with the gas accumulating faster in the inner than in the
outer region, with timescales equal to about 1 Gyr and 8 Gyr
respectively (inside-out scenario). The star formation rate is given
by the expression:
\begin{equation}
\psi(r,t)=\nu\left[{\sigma(r,t)}\over{\sigma(r_{\sun},t)}\right]^{2(k-1)}
\left[{\sigma(r,t_{Gal})}\over{\sigma(r,t)}\right]^{k-1}
\sigma_{ISM}^{k}(r,t)
\end{equation}
Where $\sigma(r,t)$ is the total mass (gas + stars) surface density
at a radius r and time t, $\sigma(r_{\sun},t)$ is the total mass
surface density in the solar region. The starformation efficiency,
$\nu$, is set to $\nu = 1$ $Gyr^{-1}$ and becomes zero when the
surface density drops below $\sigma_{th}=7$ $M_{\sun}pc^{-2}$ as
suggested by Kennicutt (1989). For the gas density exponent $k$ a
value of 1.5 has been assumed by Chiappini et al. (1997) in order to
ensure a good fit to the observational constraints of a large set of
local spirals (Kennicutt 1998). The resultant star formation history
is shown in figure~\ref{a}a for a 2 Kpc wide ring located 8 Kpc from
the Galactic centre, which represents the solar neighborhood.

The IMF adopted is a simplified two-slope approximation to the
actual Scalo (1986) IMF which is expressed by the formula:
\begin{eqnarray}
\phi_{scalo}(m) = \left\{
\begin{array}{l l}
0.19\cdot m^{-1.35} & \quad \mbox{if  $m < 2M_{\sun}$  } \\
0.24\cdot m^{-1.70} & \quad \mbox{if  $m > 2M_{\sun}$  }
\end{array} \right.
\end{eqnarray}

\subsubsection{Ellipticals}
There is a long lasting debate on the scenario for the formation of
elliptical galaxies. While it is clear that dark matter structures
build up in a hierarchical fashion from smaller to bigger ones,
there are strong hints that the much more complex physics of visible
matter, impossible to simulate from first principles, reverses, to
some level this trend. Indeed, recent observations made in the past
few years (Cowie et al. 1996; Guzman et al. 1997; Brinchmann \&
Ellis 2000; Kodama et al. 2004; Juneau et al. 2005; Bell et al.
2005; Noeske et al. 2007), add further support to previous
suggestions coming mostly from detailed studies of chemo-photometric
properties (e.g. Thomas et al. 2005 and references therein) for so
called "cosmic downsizing", that large elliptical galaxies formed
their stars at high redshift in a huge burst of star formation, and
after that evolved in a almost passive way (for a recent
comprehensive review see Renzini 2006). Indeed, a few modern
semi-analytic models (e.g. Granato et al. 2004, Bower et al. 2006)
try to obtain, within a computation of hierarchical build-up of dark
matter potential wells, a rapid and high redshift assembly of
baryons in large galaxies (which are dominated by ellipticals), in
other words breaking the hierarchy of galaxy formation and mimicking
to some extent the monolithic scenario (Somerville 2005). A
fundamental ingredient may be a proper treatment of the AGN
feedback, neglected in all previous computations. Which could make
passive these galactic system after an early (z $>$ 2) and short
phase ($\sim$ 1 Gyr) of rapid mergers of similar sub-units.

A detailed modelling of the physical mechanisms responsible for the
downsizing is clearly outside the scope of the present paper. Thus,
here we adopt and investigate the simple monolithic model proposed
in CPM08 which has been carefully calibrated against the chemistry
of ellipticals, deferring the investigation of alternative and more
complex possibilities to future work. The elliptical galaxies in
CPM08 form as a result of a rapid collapse of a homogeneous sphere
of primordial gas leading to a period of intense star-formation.
This star-formation is assumed to halt as the energy of the ISM,
heated by stellar winds and SNe explosions exceeds the binding
energy of the gas, sweeping away almost all of the residual gas. At
later times the galactic wind is maintained by type Ia SNe provided
that the thermal energy exceeds that of the binding energy of the
gas.

The star-formation rate is a Schmidt law expressed as:
\begin{equation}
 \psi=\nu G ^{k} (t)
\end{equation}
Where $G$ is the total gas mass with an exponent $k=1$. The star
formation efficiency, $\nu$, is assumed to increase with mass (as
originally postulated in Matteucci 1994) in order to reproduce the
positive correlation of [Mg/Fe] with mass (e.g. Thomas et al. 2005).
In this paper models of galaxies of three different masses have been
followed. Galaxies with a final stellar mass of $10^{10} M_{\sun}$
(the low mass galaxy described in CPM08) with a star-formation
efficiency $\nu=5$ $Gyr^{-1}$. Galaxies with a final stellar mass of
$10^{11} M_{\sun}$ (model La1 as described in CPM08) with a
star-formation efficiency $\nu=15$ $Gyr^{-1}$. Galaxies with a final
stellar mass of $10^{12} M_{\sun}$ (model Ha1 as described in CPM08)
with a star-formation efficiency $\nu=25$ $Gyr^{-1}$. The
star-formation history is shown in figure~\ref{a}b.

The IMF adopted here is a Salpeter IMF of form:
\begin{equation}
\phi_{Salp}(m) =0.17\cdot m^{-1.35}
\end{equation}
The IMF in different galactic system is a topic hotly debated in
literature. For instance, Baugh et al. (2005) showed that the only
way to reproduce the statistic of sub-mm sources, usually considered
the precursor of local ellipticals, in the context of their
"standard" semi-analytic model is to adopt an extremely top-heavy
IMF during mergers. However their model still shows discrepancies
with observed trends of $\alpha$/Fe  in local ellipticals (Nagashima
et al 2005), and predicts masses of sub-mm sources too low by more
than one order of magnitude Swinbank et al (2008).
\subsubsection{Irregulars}
Two separate formation scenarios for irregular galaxies were
proposed in CPM08. In this paper only the first scenario is
followed, in which the irregular galaxies are assumed to assemble
from the infall of protogalactic small clouds of primordial chemical
composition and to produce stars at a slower rate than spirals. This
formation scenario has been proposed to model dwarf irregular
galaxies of a magellanic type.

The star formation rate is a Schmidt law as shown in equation 3,
with $k=1$ and a low star-formation efficiency $\nu$ = 0.05
$Gyr^{-1}$. The star-formation history is shown in figure~\ref{a}c.
The IMF adopted is Salpeter as given by equation 4.
\subsection{Dust Evolution}
In addition to the gas phase the CPM08 model also follows the
evolution of several of the elements in the interstellar dust. The
elements considered are C, O, Si, Mg, Fe, S. The factors considered
are similar to those affecting the evolution of the ISM and are:
\begin{itemize}
\item The subtraction of dust elements due to star formation.
\item The production in stellar winds in low and intermediate mass
non-binary stars with masses 0.8 - 8 $M_{\sun}$.
\item The production in type Ia SNe by binary stars in
systems with combined masses 3-16 $M_{\sun}$.
\item The production in type II SNe of massive stars of masses 8 - 100 $M_{\sun}$.
\item The loss of dust in outflows due to galactic wind.
\item The destruction of dust primarily from SNe shocks.
\item The accretion of dust in dense molecular clouds (MCs).
\end{itemize}
The dust condensation efficiencies assumed in the stellar winds and
in SNe are those suggested by Dwek (1998).

The destruction rates in SNe shocks and the accretion rates in MCs
follow suggestions by McKee (1989) and Dwek (1998).
\subsubsection{Spirals}
In the paper of CPM08 only the dust properties of the solar
neighborhood are considered.

\begin{figure}
%%  % Requires \usepackage{graphicx}
 \includegraphics[type=eps,ext=.eps,read=.eps, width=8cm]{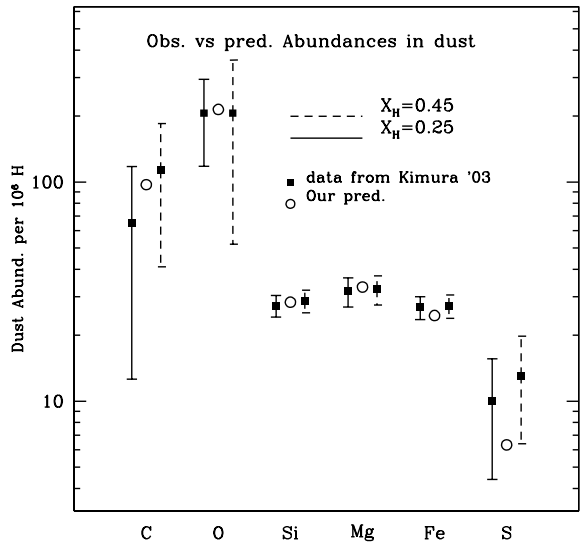}\\
 \caption{Comparison between the predicted dust abundances for the refractory elements by the CPM08 model for the solar neighbourhood (circles)and
 the elemental abundances observed in the local interstellar cloud by Kimura, Mann
\& Jessberger (2003) (squares), using two different estimates of the
ionization fraction of hydrogen} \label{b}
\end{figure}

\begin{figure}
%%  % Requires \usepackage{graphicx}
 \includegraphics[type=eps,ext=.eps,read=.eps, width=6cm, angle=90]{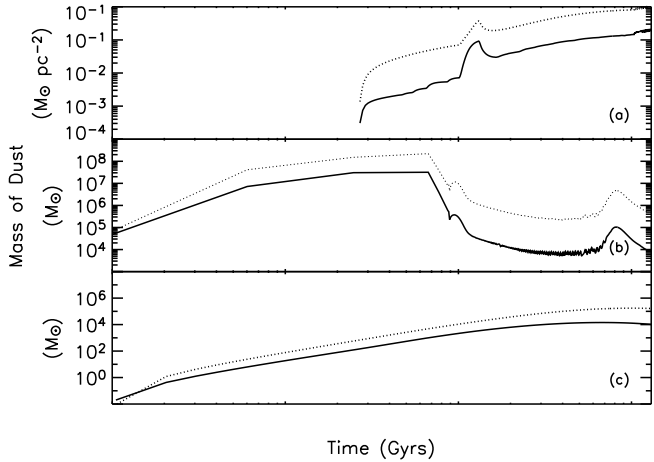}\\
 \caption{Predicted evolution of the dust masses for galaxies of different morphologies, showing evolution of carbon (solid line) and silicates (dashed line); a) spiral galaxies ; b) Elliptical galaxies; c) Irregular galaxies}
\label{c}
\end{figure}

The accretion rates are set to realistic values following Dwek
(1998) and the destruction rates are set separately for each element
and are related to their individual condensation temperatures, so
that depletion fractions of the elements in the model can reproduce
those observed in the local warm medium. The results are shown in
figure~\ref{b}.
For the purposes of this paper we are interested in following the
evolution of the mass of carbon dust particles and silicates (dust
particles composed of O, Mg, Si, S, Ca, Fe). The evolution of these
particles are shown in figure~\ref{c}a.

The solar neighborhood is believed to be representative of the
Galaxy as a whole. Hence, for this paper, the properties calculated
for the solar neighborhood will be used as global properties for the
whole galaxy.
\subsubsection{Ellipticals}
For the initial starburst phase the same efficiencies for accretion
and destruction as in the solar neighborhood are assumed. However,
after the onset of the galactic winds elliptical galaxies will be
largely devoid of cold gas. Therefore it is assumed that no
accretion can occur and a much lower value for the destruction
efficiency is used, as derived by McKee (1989), for a hot and
rarefied medium. Also an additional destruction term is included to
treat the thermal sputtering which is thought to be the dominant
source of destruction in hot plasmas. Following Itoh (1989) we
assume that in a $\sim$ 1 keV plasma nearly 90 per cent of the dust
grains will evaporate by sputtering in $\tau_{destr_{sput},i}\sim
10^{5}/n_e$ $(yr cm^{-3})$ where the electron density has been
self-consistently evaluated at each timestep. The resultant dust
evolution is shown in figure~\ref{c}b.
\subsubsection{Irregulars}
For irregular galaxies the destruction value for the solar
neighborhood is used. However, since the observed molecular H
content in dwarf irregular galaxies is very small, with
molecular-to-atomic gas fractions of $\sim$ 10 per cent or lower
(Lisenfeld \& Ferrara 1998, Clayton et al. 1996) it is assumed that
no accretion can occur. The resultant dust evolution is shown in
figure~\ref{c}c .
\section{The Stellar Population and Dust Model}
The far-UV to radio SEDs of the models discussed in section 2 will
be calculated using the GRASIL code which follows the evolution of
the stellar population, constructing a SED at any given time, first
taking into account the extinction and emission by dust. All the
details are given in S98, and only a summary of the main features
are presented here.

Though during certain phases of galaxy evolution AGN activity may
substantially contribute in shaping the observed SED of the galaxy,
both providing an additional point-like source in the center of the
galaxy and contributing to the heating of galactic dust, in the
present paper we do not consider this complication, which we
postpone to future investigation. This is justified by the fact that
QSO lifetimes are commonly estimated to be less than 100 Myrs (e.g.
McLure \& Dunlop 2002, Marconi et al 2004, Shankar et al 2005,
Hopkins \& Hernquist 2008), so that during the majority of the life
of a galaxy any effect will be negligible.
\subsection{Stellar Population Model}
The single stellar populations included in GRASIL are based on the
Padova stellar models and cover a large range in ages and
metallicity. The SSPs include asymptotic giant branch (AGB)
isochrones, which incorporate a treatment of the dusty envelope
around these stars, and have been calibrated on the most recent data
for the V-K colours of the large magellanic cloud's star clusters
(see Marigo et al. 2008; Bressan et al. 1998). The spectral
synthesis technique consists in summing up the spectra of each
stellar population provided by a simple stellar population (SSP) of
appropriate age and metallicity (Z), weighted by the star-formation
rate (SFR) ($\Psi$) at the time of the stars birth (Bressan et al.
1994):
\begin{equation}
F_{\lambda}(t_{G}) =
\int\limits_{0}^{t_{G}}SSP_{\lambda}(t_G-t,Z(t))\times\Psi(t)dt
\end{equation}
Where $t_G$ is the age of the galaxy and $t$ is the birth age of an
individual SSP (see also below the treatment of age dependent
attenuation). The SFR and the metallicity values are those
calculated by the chemical evolution model (see section 2.1).
\subsection{Dust Model}
GRASIL then calculates the radiative transfer of the starlight, the
heating of the dust grains and the emission from these grains ,with
a self-consistent calculation of grain temperatures, for an assumed
geometrical distribution of the stars and dust and a specific grain
model.
\begin{figure}
%% Requires \usepackage{graphicx}
 \includegraphics[type=eps,ext=.eps,read=.eps, height=9cm,angle=270]{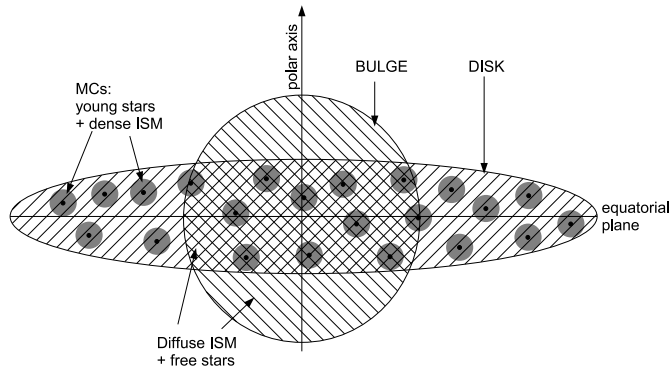}\\
 \caption{Sketch of geometry of stars and dust in the GRASIL model for the case of a galaxy with both a disc and a bulge.} \label{d}
\end{figure}

The stars can be modeled in two main components: the bulge and the
disc, the latter not being present when modeling elliptical
galaxies. The geometry is shown in figure~\ref{d}. The spherical
bulge has an analytic King profile, $\rho\propto(r^2+r^2_c)^{-3/2}$
extended up until the tidal radius $r_t = 10^{2.2}r_c$, where the
scalelength $r_c$ is a free parameter. The disc is described by a
radially and vertically exponential profile, $\rho\propto
exp(-R/R_d)exp(-|z|/z_d))$ truncated at $r_t=6R_d$ where the radial
scalelength $R_d$ and vertical scalelength $z_d$ are free
parameters.

The dust component has the same disc and bulge profiles as the
stellar component but can have scalelengths independent of the
stellar component. Within these profiles the gas and dust is further
split into two main components: dense MCs and diffuse cirrus. The
ratio of the total mass in dust and gas located in the MCs to the
total gas mass is a free parameter $f_{mc}$. When a galaxy is
composed of only a bulge or a disc the MCs are dispersed evenly
throughout the galaxy, however when a galaxy is modeled with both a
disc and a bulge, as shown in figure~\ref{d} the MCs are located
solely in the disc. Stars are assumed to form inside the MCs and
progressively escape them on a timescale $t_{esc}$. Specifically,
the fraction of stars inside clouds at time $t$ after they formed is
assumed to be:
\begin{eqnarray}
F(t)=\left\{
        \begin{array}{l l}
1                    &  (t<t_{esc})           \\
2 -{{t}\over{t_{esc}}} &  (t_{esc}<t<2t_{esc}) \\
0                    &  (t>2t_{esc})
        \end{array}
\right.
\end{eqnarray}
Where $t_{esc}$ is a free parameter of the model. The radii and mass
of the MCs are also free parameters of GRASIL, though the results
actually depend only on the combination $M_{mc}/R_{mc}^{2}$ which
determines together with the dust-to-gas ratio, the optical depth of
the clouds (see S98), where $M_{mc}$ is the mass of a MC and
$R_{mc}$ is the radius of a MC.

The dust is assumed to consist of a mixture of carbonaceous and
silicate grains and polycyclic aromatic hydrocarbons (PAH). The
carbonaceous particles are assumed to have the optical properties of
randomly-orientated graphite spheres. The silicate grains are
assumed to be amorphous silicates. The optical properties of
silicate and graphite grains are taken from Laor \& Draine (1993),
the optical properties of PAH molecules from Draine \& Li  (2007).
The size distribution used in this paper for our fiducial models are
those chosen by S98 to match the extinction and emissivity curves of
the local ISM. The size distribution for both the carbon and
silicate components are described by a double power law specified by
equation 7.
\begin{eqnarray}
\frac{dn_i}{da} = \left\{
 \begin{array}{l l}
     A_i n_{_H} a^{\beta_1} & a_b < a < a_{max}\\
     A_i n_{_H} a^{\beta_1 - \beta_2} & a_{min} < a < a_b
   \end{array}
\right.
\end{eqnarray}
Where $dn_i$ is the number density of grains of type i with radii in
the interval  [a, a + da], $n_H$ is the number density of H nuclei
and $A_i$ is the atomic abundance of type i relative to hydrogen.
The size ranges  $(a_{min})$, $(a_{b})$ and $(a_{max})$ of the dust
grains and the exponents $\beta_1$ and $\beta_2$ are free parameters
with values shown in table 1.
\begin{table}
\caption{Parameters for size distribution of the dust component in
different environments (see equation 7). The MW size distribution is
that derived by S98 to match observations from the MW. The SMC and
QSO SDSSJ1048+46 size distributions are calculated in this paper to
match observed extinction curves.}
\begin{tabular}{| c | c c c }
    \hline
                    & MW (S98) & SMC & QSO SDSSJ1048+46 \\
    \hline
    $\mathbf{Carbon}$   &         &         &        \\[5pt]
    $a_{min}$ $(\AA)$      & 8       & 5       & ...   \\[5pt]
    $a_{b}$ $(\AA)$        & 50      & 25      & 400   \\[5pt]
    $a_{max}$ $(\AA)$      & 2500    & 1000    & 700   \\[5pt]
    $\beta_{1}$        & -3.5    & -3.0    & -3.5  \\[5pt]
    $\beta_{2}$        & -4.0    & -4.5    & ...      \\[5pt]
    $\mathbf{Silicate}$ &         &         &         \\[5pt]
    $a_{min}$ $(\AA)$      & ...     & ...     &...          \\[5pt]
    $a_{b}$ $(\AA)$        & 50      & 50      & 5       \\[5pt]
    $a_{max}$ $(\AA)$      & 2500    & 10000   & 10000       \\[5pt]
    $\beta_{1}$        & -3.5    & -3.9    & -3.5       \\[5pt]
    $\beta_{2}$         & ...     &  ...    & ...      \\[5pt]
\hline
\end{tabular}\\
\end{table}
In this work the total mass of dust and also the fraction of the
graphite component to the silicate one is that calculated by the
CPM08 chemical evolution model.

The luminosities of the different stellar components (the stars in
the bulge, in the disc and young stars still in clouds) are
calculated using the population synthesis model described. The
GRASIL code then calculates the radiative transfer of the starlight
through the specified dust distribution. For the MCs a full
radiative transfer calculation is performed, however for the diffuse
dust the effects of scattering are only approximated by assuming an
effective optical depth related to the true absorption and
scattering optical depths by:
$\tau_{eff}=[\tau_{abs}\,(\tau_{abs}+\tau_{sca})]^{1/2}$. Thus
GRASIL calculates for each point inside and outside the galaxy the
dust attenuated stellar radiation field. Using these values it is
then possible to calculate for each point inside the galaxy the
absorption of radiation, thermal balance if in thermal equilibrium
or thermal fluctuations if not, and hence, re-emission for each
grain composition and size. Combining the contributions from the
attenuated starlight and from the dust re-emission a SED for the
galaxy is calculated which depends on the angle the galaxy is viewed
from.
\subsection{Choice of parameters}
The value of the GRASIL parameters chosen for the different galaxies
are listed in table 2. The reasons for the choices made are given
below:
\begin{table}
\caption{Adopted Values for the GRASIL parameters, - for the
elliptical models the values outside the brackets are for the
starbursting phase, the values in the brackets are for the passively
evolving phase}
\begin{tabular}{| c | c c c c c}
    \hline
                  & {Spiral}& \multicolumn{2}{c}{Elliptical} &{Irregular}  \\
                  &          & $10^{11} M_{\sun}$&$10^{12}
M_{\sun}$&    \\
    \hline
    $t_{esc}$ $^1$ & 2             & 250 (0)       &  250 (0)       &  2      \\[5pt]
    $f_{mc}$ $^2$        & 0.5           & 0.5 (0)        &  0.5 (0)       &  0.1    \\[5pt]
    $r_c^{*}$ $^3$ & ...       & 0.21 (0.21)   & 0.68 (0.68)    & ...     \\[5pt]
    $r_d^{*}$ $^4$ & 2.35         & ... (...)     & ... (...)    &  1      \\[5pt]
    $z_d^{*}$ $^5$ & 0.14        & ... (...)     &  ... (...)   &  0.5    \\[5pt]
    $r_c^{d}$ $^6$ & ...        & 0.21 (0.42)   & 0.68 (1.36)    & ...     \\[5pt]
    $r_d^{d}$ $^7$ & 5           & ... (...)    & ... (...)    &  1      \\[5pt]
    $z_d^{d}$ $^8$ & 0.1       & ... (...)   & ... (...)    &  0.5    \\[5pt]
    $M_{mc}$ $^9$ &$10^6 $  & $10^6$ ($10^6$)& $10^6$ ($10^6$) &$10^6$   \\[5pt]
    $R_{mc}$ $^{10}$       &16            &  16 (16)      &  16 (16)       &40       \\[5pt]
\hline
\end{tabular}
\\
$^1$ Timescale for the evaporation of MCs, in $Myrs$. $^2$
Fractionary gas content in the MCs. $^3$ Radial scalelength of the
stellar component in the bulge, in $Kpc$. $^4$ Radial scalelength of
the stellar component in the disc, in $Kpc$. $^5$ Vertical
scalelength of the dust component in the disc, in $Kpc$. $^6$ Radial
scalelength of the dust component in the bulge, in $Kpc$. $^7$
Radial scalelength of the dust component in the disc, in $Kpc$. $^8$
Vertical scalelength of the dust component in the disc, in $Kpc$.
$^9$ Total gas mass in each Mc, in $M_{\sun}$. $^{10}$ Radius of
each MC, in $pc$
\end{table}
\subsubsection{Escape timescale ($t_{esc}$)}
This is a very important parameter controlling how long the young
stars remain in their birth clouds. A value of 2 Myrs for spiral and
irregular galaxies is chosen following Granato et al. (2000). S98
found that larger values were needed to fit local starbursts:
$t_{esc}=20-60$ Myrs, to reflect the dense dusty environments found.
In a similar way one would expect much higher escape timescales for
the early starbursting phase of elliptical galaxies. The rates of
star-formation however predicted by the CPM08 model far exceed those
observed in local starbursting galaxies and as a result the escape
timescale could reasonably be expected to be longer still. This
value remains a free parameter of the model since the precise dust
condition present in SCUBA galaxies are unknown. In this work it has
been set to 250 Myrs, a value found to provide good fit to SCUBA
galaxies at high redshift (see section 5.3.1). After the onset of
the galactic winds in the elliptical galaxy all of the cold gas is
blown out of the galaxy and MCs cannot form, as a result star
formation is stopped and consequently $t_{esc}=0$.
\subsubsection{Optical depth of MCs ($M_{mc},R_{mc}$)}
Combined these two parameters control the optical depth of the MCs
according to the ratio $M_{mc}/R_{mc}^{2}$, where $M_{mc}$ is the
mass of a MC and $R_{mc}$ is the radius of a MC. Following Granato
et al. (2000) values of $M_{mc}=10^6 M_{\sun}$ and $R_{mc}$=16 pc
are chosen for the spiral and elliptical galaxies, values which have
been found to give good SED fits and are consistent with
observations.

GRASIL has not been previously used to model irregular galaxies. In
order to fit the SED in the MIR MCs with a lower optical depth had
to used in this work (see section \emph{5}).
\subsubsection{Ratio of gas mass in MCs to total gas mass ($f_{mc}$)}
For spiral galaxies we choose a value of $f_{mc}$ = 0.5 for this
parameter which agrees well with the values used to fit local
spirals in S98.

The passively evolving late phase in elliptical galaxies will have
no MCs so $f_{mc} = 0$. A value of 0.5 is adopted for the
starbursting stage.

Lower value for the molecular-to-atomic gas fraction are observed in
irregular galaxies than in spiral galaxies (see section 2.2.3) and
since most of the hydrogen in MCs is in $H_{2}$, while most of the
diffuse cirrus gas will be atomic $H_{1}$ this lower value suggests
a lower value of $f_{mc}$. Consequently a value of $f_{mc} = 0.1$
has been chosen.
\subsubsection{Geometry ($r_c^{*}$, $r_d^{*}$, $z_d^{*}$, $r_c^{d}$, $r_d^{d}$, $z_d^{d}$)}
The spiral galaxy is modeled as a pure disc for both the stellar and
the dust components with scalelengths, ($r_d^{*}$ and $z_d^{*}$ for
the stellar component and $r_d^{d}$ and $z_d^{d}$ for the dust
component) set following the results of Misiriotis et al. (2006),
who used COBE/DIRBE maps and COBE/FIRAS spectra to constrain a model
for the spatial distribution of the dust, the stars and the gas in
the Milky Way.

The elliptical galaxies are modeled using a pure bulge for both the
stellar and the dust components. The elliptical stellar scalelength,
$r_c^{*}$ is given by the chemical evolution model of CPM08. S98
found that in order to obtain a good fit to a template constructed
from local elliptical galaxies a dust scalelength, $r_c^{d}$, much
larger than the stellar scalelength was required. This conclusion
was also reached by Panuzzo et al. (2007) who found it necessary to
have dust scalengths almost double the stellar ones in order to fit
a late type galaxy of the Virgo cluster. Spatially extended dust has
also been observationally detected by Temi, Brighenti $\&$ Mathews
(2007), who have detected spatially extended dust in two optically
normal galaxies NGC 5044 and NGC 4636. Although it should be noted
that Temi et al. (2007) concluded that due to the short sputtering
lifetime for extended dust $\sim$ 7 Gyrs and no evidence for a
recent merger that the most likely source of this dust was due to
galactic winds powered by a central AGN, whereas in the CPM08 model
the proposed mechanism for the galactic wind is different, with SN1a
and stars alone able to power the wind for several Gyrs (see Pipino
et al. 2005). Therefore due to observational evidence and also
typical values used in previous studies, in the passively evolving
phase, we have set the dust scale-length to be twice the stellar
scale-length. For the early star-bursting phase the dust is likely
to be more concentrated, so the dust scale-length is set equal to
the stellar one.

The irregular galaxy is modeled as a pure disc for both the stellar
and the dust components. Observations for the dust distribution in
irregular galaxies are still largely inconclusive so we have set the
dust scale-length equal to the stellar one.
\subsection{Generating SEDs}
In order for GRASIL to be able to calculate the intrinsic stellar
SED of a galaxy the stellar population code requires a star
formation history and a metallicity evolution which are generated by
the chemical evolution code, and an IMF which is set to that used in
the chemical evolution code. To calculate the dust effected SED
GRASIL requires a total mass in dust (in this work the dust-to-gas
ratio will be the quantity discussed) and chemical composition of
dust (the carbon over silicate ratio) which are set by the dust
evolution model the other geometrical parameters are then set to the
values described in section 3.3 and the grain size distribution to
that derived in S98.

For comparison SEDs are also calculated for galaxies with the
chemical evolution described by the CPM08 model but without the
detailed dust evolution properties. Instead two common assumptions
will be adopted.
\begin{itemize}
\item That the ratio of the dust mass to gas mass is proportional to the
metallicity, through a constant normalised to the Milky Way, using
the value 0.008 calculated from observations for the dust to gas
mass ratio as discussed in CPM08. So \emph{dust mass} $\propto
metallicity \times$ \emph{gas mass}.
\item That the chemical composition does not evolve. So that the fraction of carbon to silicate will be set to the
value 0.19 for all times and morphologies, a theoretical value found
by CPM08 for the ratio of C/Si at 13 Gyr in the solar neighborhood
model.
\end{itemize}
In addition, for the irregular galaxy and for the starbursting phase
in elliptical galaxies, the effect on the SED of releasing a third
assumption, made routinely in other works as well as for the
fiducial models in this paper, namely that the dust size grain
distribution does not evolve and remains constantly identical to
that of the Milky Way is investigated.

For clarity when the full dust treatment is used as described in the
CPM08 paper it will be referred to as the full CPM08 model. When the
two assumptions outlined above are adopted it will be referred to as
the simple dust model. The difference between the two models will be
referred to as the \emph{disparity}. This would be the error
introduced by the adoption of the simplifications if one assumes
that the fiducial model is correct. It is only a theoretical error.
\subsection{Treatment of PAH molecules}
The abundance of PAHs is calculated from the chemical composition of
the dust predicted by the CPM08 model, specifically their abundance
is proportional to the total abundance of carbon molecules in the
dusty component of the ISM. The exact treatment of the absorption
and emission processes for the PAH bands are then calculated
following the same procedure described in Silva et al. (1998) and
Vega et al. (2005) but updated with the most up-to-date
cross-sections outlined in Draine \& Li (2007) calculated using new
laboratory data and spectroscopic data observed by the \emph{spitzer
space telescope}

Although PAH emissions have been found to be a ubiquitous feature of
both the Milky Way and external galaxies it is worth noting that the
strengths of the PAH bands have been found to vary between galaxies,
with low metallicity galaxies and galaxies hosting AGNs in
particular being found to be deficient in PAH emission. This has led
several authors to speculate that the abundances of PAHs could be
linked to the particular environmental conditions present in their
host galaxy. For example it it thought that they could be
preferentially destroyed by strong UV radiation (e.g. Madden et al.
2006) and, could be more susceptible to SNe shocks (e.g. O'Halloran
et al. 2006). This suggests that the evolution of PAHs may not
necessarily be identical to that of the larger carbon grains. This
effect has been partly accounted for by following the treatment of
PAHs outlined in Vega et al. (2005). Where in order to agree with
observations from around hot stars and H$_{II}$ regions, from within
our galaxy, the abundance of PAHs in the MCs within GRASIL, where
the UV radiation will be strongest, had to be suppressed by a large
factor ($\approx$ 1000 with respect to the cirrus component). Using
such a prescription it was shown that GRASIL could reproduce the MIR
spectrum of local star-forming galaxies.

Since in this work the evolution of the PAH dust population has not
been specifically followed a detailed analysis of the spectrum in
the MIR will not be possible. However linking the abundance of the
PAHs to both the environment, by suppressing the PAH abundance in
MCs, and to the chemical composition of the dust should be a
reasonable approximation and should be sufficient to highlight broad
trends in the emission from this region.
\subsection{Comparison to Observations}
\subsubsection{Low Redshift}
The theoretical SEDs are compared to observed nearby galaxies of
similar morphological type. The data sample used is that described
in Dale et al. (2007) which presents SEDs for the 75 galaxies of the
SINGS sample selected to span a wide range of morphologies,
luminosities and IR to optical ratios. The data set comprises of
galaxies at an average distance of $\sim 10$ Mpcs and contains SEDs
from UV to radio wavelengths and includes :
\begin{itemize}
  \item UV data from GALEX
  \item optical fluxes from the RC3 catalogue (de
   Vaucouleurs 1991).
  \item IR data from 2MASS, ISO, IRAS, SPITZER
  \item Sub-mm data from SCUBA
  \item Radio data - from literature including the New VLA Sky survey
\end{itemize}
All the galaxies of a particular morphological type are normalised
to a common value in the \emph{K}-band.

Our spiral model is designed to reproduce observations of the Milky
Way, hence, to represent this morphological type, all spirals of
type SBb to SBc and SAb to SAc in the SINGs data set were chosen,
numbering twenty four. Our irregular galaxy is designed to reproduce
observables of magellanic type irregular galaxies therefore the ten
galaxies of type Im will be used. Our elliptical galaxy model is
designed to reproduce the observables of passively evolving
elliptical galaxies. Of the six elliptical galaxies in the SINGS
sample two NGC 855 and NGC 3265 show very bright UV colours in
addition to srtong IR emission probably indicating ongoing moderate
starformation. These two galaxies were identified as starforming
galaxies by Dale et al. (2007) and were therefore not used in this
work.
\subsubsection{High Redshift}
In order for the elliptical galaxy formation scenario presented in
this work to be correct, one would expect to find observed galaxies
at high redshift with SEDs similar to those presented in this work.
It is therefore essential to check that our young elliptical models
are representative of the birth of typical elliptical galaxies at
high redshift, by comparing the theoretical models to observations.
However, for each of the masses of galaxy, only one star formation
history has been followed with a fixed set of values for all of the
parameters, so the models could not be expected to account for all
high redshift objects. Also due to the sheer variability of high
redshift galaxies the construction of an 'average' high redshift SED
is not possible. Consequentially we compare our models to two
different high redshift data sets (details of which can be found in
the relevant sections), each thought to be representative of a
different stage of the young galaxies evolution, to check that a
reasonable proportion of the observed galaxies have SEDs similar to
that predicted by our models.

A chi-squared fitting technique was employed to calculate the age
and redshift of the model which gives the best fit to the observed
points. Theoretical SEDs for the elliptical galaxy in its
starbursting and post-starbursting phases were generated at several
time steps, from 0.05 Gyrs to 2 Gyrs. Each of these SEDs is then
normalised and compared to the observed spectra of the galaxy. Since
neither of the data sets have spectroscopic redshifts the
theoretical SEDs at all the time steps are compared to the observed
SED for a range of redshift values and then the age and redshift
which corresponds to the best fit is selected using a chi-squared
minimisation technique. Due to the inherent uncertainty from using
such a limited template set, in order to avoid spurious results the
possible redshift range will be limited to within 30 per cent of the
photometric redshift values calculated previously by other authors,
using more sophisticated fitting procedures.
\section{Results}
\subsection{Properties of Spiral Galaxies}
\begin{figure}
%%  % Requires \usepackage{graphicx}
 \includegraphics[type=eps,ext=.eps,read=.eps, width=6cm,angle=90]{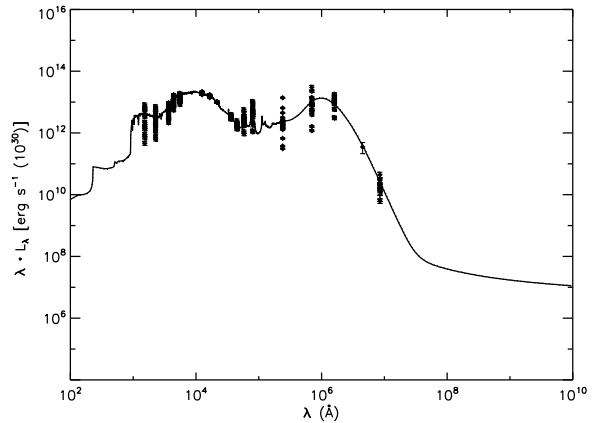}\\
 \caption{Comparison of theoretical spiral model at 12 Gyrs with the SEDs
from all spiral galaxies of morphological type ranging from SBb to
SBc and SAb to SAc in the SINGS sample, normalised to the k band.
Solid line - Full CPM08 model. Dotted line - simple dust model}
\label{e}
\end{figure}

Figure~\ref{e} shows the results for the theoretical SEDs generated
at 12 Gyrs compared to the mean SED of spiral galaxies of types SBb
to SBc and SAb to SAc from the SINGS sample. The theoretical model
shows a good correlation with the observed spiral galaxies. This
suggests that the assumption that the global properties of the
galaxy can be represented by the chemical and dust evolution model
of the solar neighbourhood at late times is a good one.
\begin{figure}
 \includegraphics[type=eps,ext=.eps,read=.eps, width=6cm,angle=90]{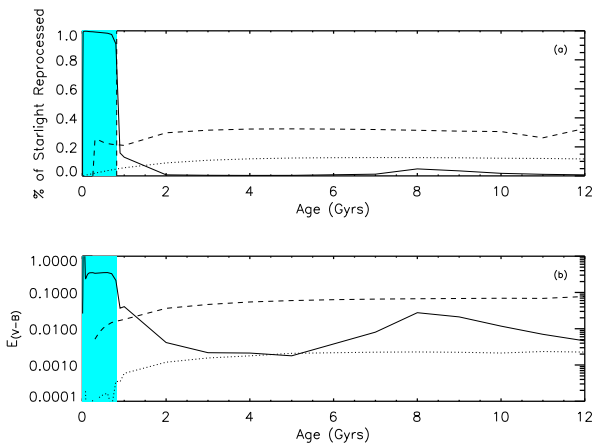}\\
 \caption{a) Fraction of stellar radiation dust reprocessed:
 Solid line - elliptical, dotted line - irregular, dashed line - spiral b) Evolution of extinction
coefficient:  Solid line - elliptical, dotted line - irregular,
dashed line - spiral. The blue shaded area represents the age before
the galactic wind starts in the elliptical model during which it is
undergoing a period of extreme starformation} \label{f}
\end{figure}

\begin{figure*}
 \includegraphics[type=eps,ext=.eps,read=.eps, width=13cm,angle=90]{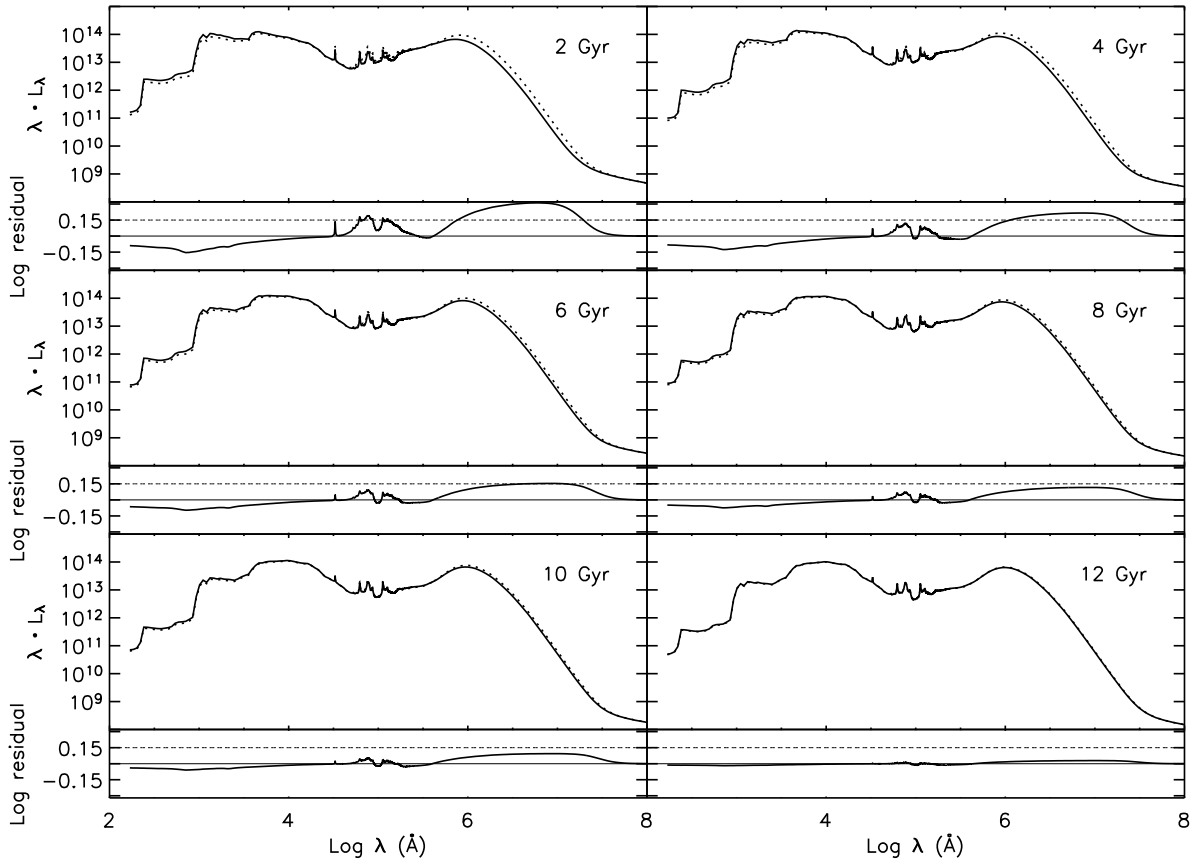}\\
 \caption{The evolution of the SED of the spiral galaxy. The larger six plots shows the SEDs at different times in the
evolution of the galaxy (with units erg $s^{-1} (10^{30})$). The
solid line shows the SEDs generated using the full CPM08 model, the
dashed line the evolution adopting the simple dust model. The lower
panels displays the log residual between the two models.} \label{g}
\end{figure*}

The evolution of the SED of spiral galaxies is shown in
figure~\ref{g}. Due to the mainly quiescent slow evolution of the
spiral galaxies the SEDs do not show large changes throughout their
lifetime and there is little variation in the percentage of
starlight reprocessed by the galaxy which for ages greater than 1
Gyr remains fairly constant at a value of 30 per cent (see
figure~\ref{f}).
\begin{figure}
 \includegraphics[type=eps,ext=.eps,read=.eps, width=6cm,angle=90]{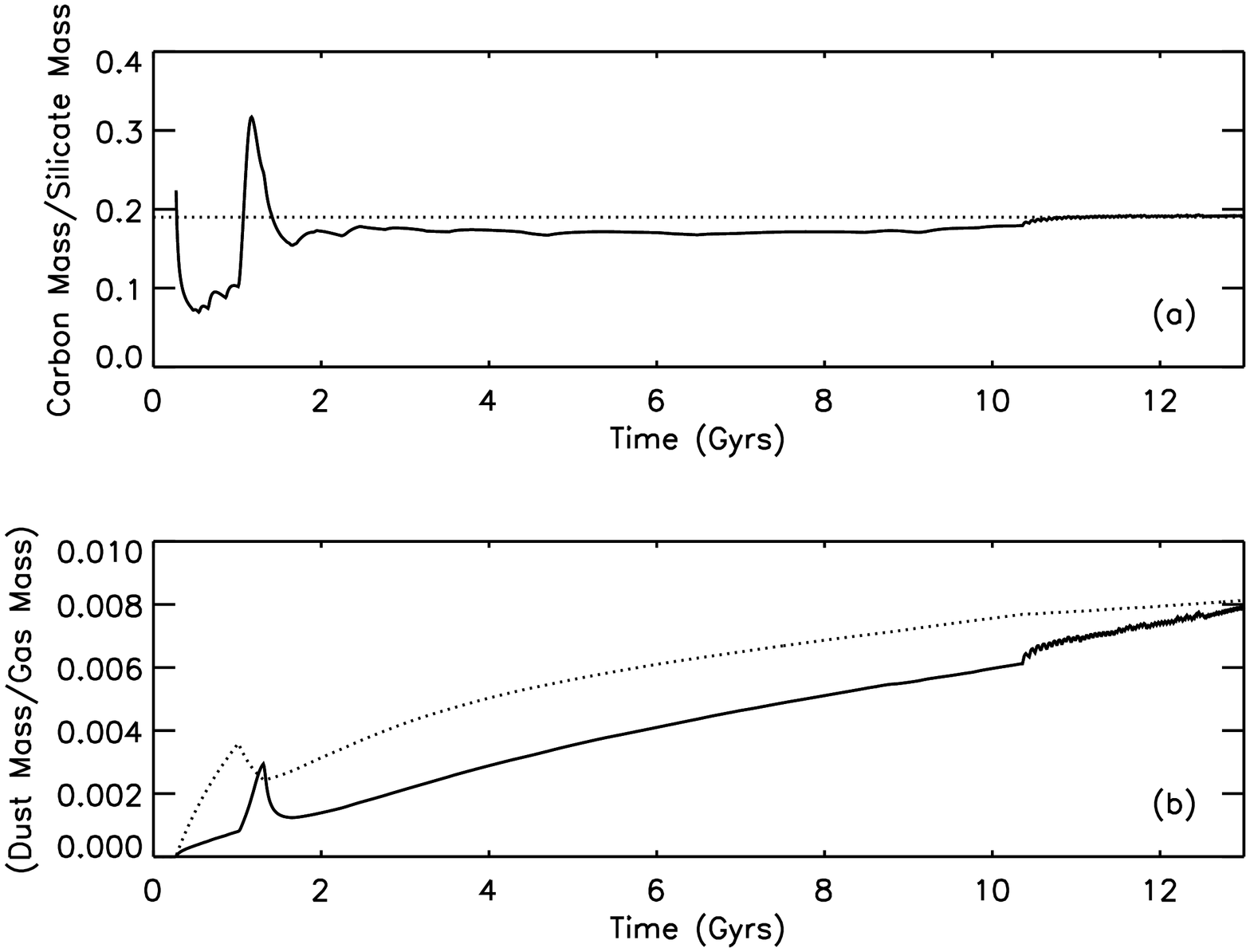}\\
 \caption{Spiral Galaxy: (a) Comparison between the value of the C/Si calculated by the full CPM08 ratio (solid line) with
the constant value assumed if the chemical abundance remains
identical to the present value for the solar neighbourhood (dotted
line). (b) Comparison of the dust-to-gas ratio value calculated by
the full CPM08 model (solid line) with the expected value if a
dependence on the metallicity is assumed (dotted line).} \label{hh}
\end{figure}

\begin{figure}
%%  % Requires \usepackage{graphicx
 \includegraphics[type=eps,ext=.eps,read=.eps, width=6cm,angle=90]{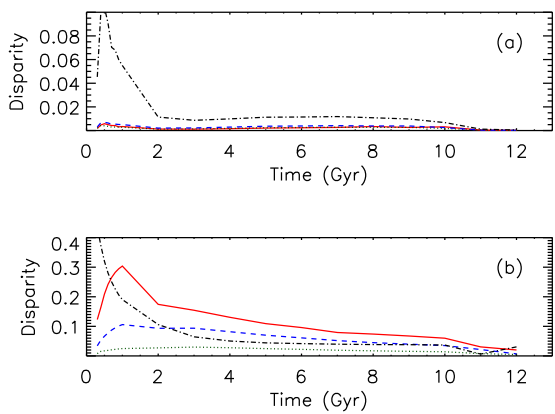}\\
 \caption{The residuals between the SEDs generated by the full spiral CPM08 model and SEDs generated by using the simplifications that
(a) the chemical composition remains equal to the Milky Way values
at all times, (b)the dust-to-gas ratio scales with the metallicity.
Blue dashed line - UV, green dotted line - optical/NIR, Black
dot/dashed line MIR, Red solid line - FIR. The disparity is given as
the absolute mean log value of the residuals between the two models
in the specified wavelength region} \label{i}
\end{figure}

Also included in figure~\ref{g} are the SEDs generated for the same
chemical evolution model but adopting the two dust assumptions.
Since the simple dust assumptions are normalised to the full CPM08
model for the solar neighbourhood at 13 Gyrs the SEDs at 12 Gyrs are
still in very close agreement. As the age of the galaxy decreases
the residuals increase (figure~\ref{i}) as the assumptions made for
the dust-to-gas ratio and carbon over silicate ratios diverge from
our theoretically calculated values.

Of the two simplifications the disparity introduced by assuming a
dust-to-gas ratio dominate the residuals. Assuming a constant Milky
Way chemical composition of the dust has little effect on the SED
because after initial fluctuations during the first 2 Gyrs this
ratio remains fairly constant, throughout the rest of its lifetime
(figure~\ref{hh}a). Hence making the simplification that the
chemical composition does not evolve with time will only lead to
small differences in the SED (figure~\ref{i}a) for the majority of
the spiral galaxies life. At early times the different chemical
composition calculated mainly effects the MIR part of the SED due to
the resulting differences in the abundance of PAHs.

The assumed dust-to-gas ratio however leads to a more significant
disparity particularly when the galaxy is younger (figure~\ref{i}b)
since our model predicts a relationship between the dust-to-gas
ratio and the metallicity which is not entirely linear
(figure~\ref{hh}b). Therefore adopting this assumption will
introduce disparities which largely increase as the age of the
galaxy decreases. This assumption will result in too much dust in
the galaxy and this will lead to an excess of reprocessing. The
disparity will be most significant in the FIR part of the spectrum
(figure~\ref{i}b).

In this paper only the chemical evolution in the solar neighbourhood
has been followed and the SED of the whole galaxy has been
constructed by globalising the results. Although this has been shown
to be a reasonable assumption for old spiral galaxies in the local
universe, since the SEDs generated are representative of
observations, it is unlikely to be valid at higher redshifts. The
chemical evolution has two infall episodes; first the halo and the
thick disc forms followed by the second phase in which the thin disc
forms. So at early times the halo and thick disc will dominate the
galaxies properties and the global properties could differ greatly
from that of the solar neighborhood. The results for spiral for very
young galaxies (ages $\leq$ 1 Gyr) should therefore be taken with
caution.

The CPM08 model and indeed all complete dust evolution models
currently written only deal with the total mass of different
elements contained in grains. As a result no information is
contained regarding how these elements are combined to form grains
and what size distribution these grains should have. Instead a
composition and size distribution for the dust grains must be
assumed. In previous works using GRASIL it has been usual to use the
grain distribution population suggested in S98. In that paper both
the observed extinction curves from the diffuse medium within the
Milky Way and the observed galactic emission from dust was
reproduced using a combination of graphite and silicate grains and
PAH molecules.

Our spiral chemical and dust evolution model is designed to match
observed properties of the solar neighbourhood and as a result it
should also be capable of reproducing the observed galactic
extinction and emission. Our model however predicts a slightly
different ratio for the dust-to-gas ratio as well as a different
graphite to silicate ratio than that used in S98. Therefore if we
use the S98 size distribution in combination with the theoretical
quantities of dust we have calculated the resultant extinction and
emission properties will differ slightly from those observed.

No adequate dust model was found in the literature which could
satisfactorily reproduce the observed extinction curves with the low
value of 0.19 for the ratio of  carbon to Silicon predicted by our
model (for example see Zubko et al. 2004). Hence, for simplicity the
S98 dust composition already included within the GRASIL model was
used. A likely possibility is that existing dust models are tuned on
low density environments (cirrus) while our C/Si value should refer
to a galactic average.
\subsection{Properties of Irregular Galaxies}
Figure~\ref{j} shows the results for the theoretical SEDs generated
at 12 Gyrs compared to the SEDs for all the magellanic type
irregular galaxies of the SINGS sample. Although the irregular
galaxies show some scatter in their SEDs the theoretical model
manages to fit the average properties remarkably well. Also shown in
the figure is the SED generated if the two dust simplifications are
used. By using these simplifications quite a clear discrepancy
between the two SEDs is introduced particularly in the IR and UV.
Although still largely consistent with the observed SEDs there is a
strong hint that the simplified dust treatment is leading to more
dust reprocessing than that observed in the average galaxy in the
sample. A common measure of the quantity of dust reprocessing in a
galaxy is the observed IR/UV ratio (e.g. Dale 2007). Using the same
formalism as Dale et al. (2007) and equation 4 from Dale \& Helou
(2002) the median Total IR to UV ratio was calculated for the
irregular magellanic galaxies in the sample to be 0.22. This agrees
well with the value of 0.21 calculated from the results of the full
CPM08 model. For the simple dust model however a value of 0.67 was
calculated implying that to much energy has been absorbed in the UV
and emitted in the IR.

For the most part the disparity between the SED generated using the
full dust treatment and that generated using the simplifications can
be attributed to the assumption that the dust-to-gas ratio scales
with metallicity (figure~\ref{k}b). As figure~\ref{m}b shows the
theoretical values calculated by our model are much lower than those
calculated from the  Milky Way scaling relation, by approximately a
factor of 6. This much lower dust to gas ratio can be mainly
attributed to the absence of accretion in our irregular galaxy model
which will results in smaller quantities of dust forming than would
be expected following the common Milky Way metallicity scaling
relation, where dust accretion is prevalent. Such a finding is
consistent with the work of Galliano et al. (2003 \& 2005) where
dust masses were estimated for 4 local dwarf galaxies and for two
(NGC 1569 and He 2-10) values for the dust-to-gas ratio lower by a
factors 4-7 and 2-10 than expected were found and Hunt et al. (2005)
and Walter et al. (2007) who estimated the dust masses in local
dwarf galaxies and found values consistently smaller than expected.

\begin{figure*}
 \includegraphics[type=eps,ext=.eps,read=.eps, width=13cm,angle=90]{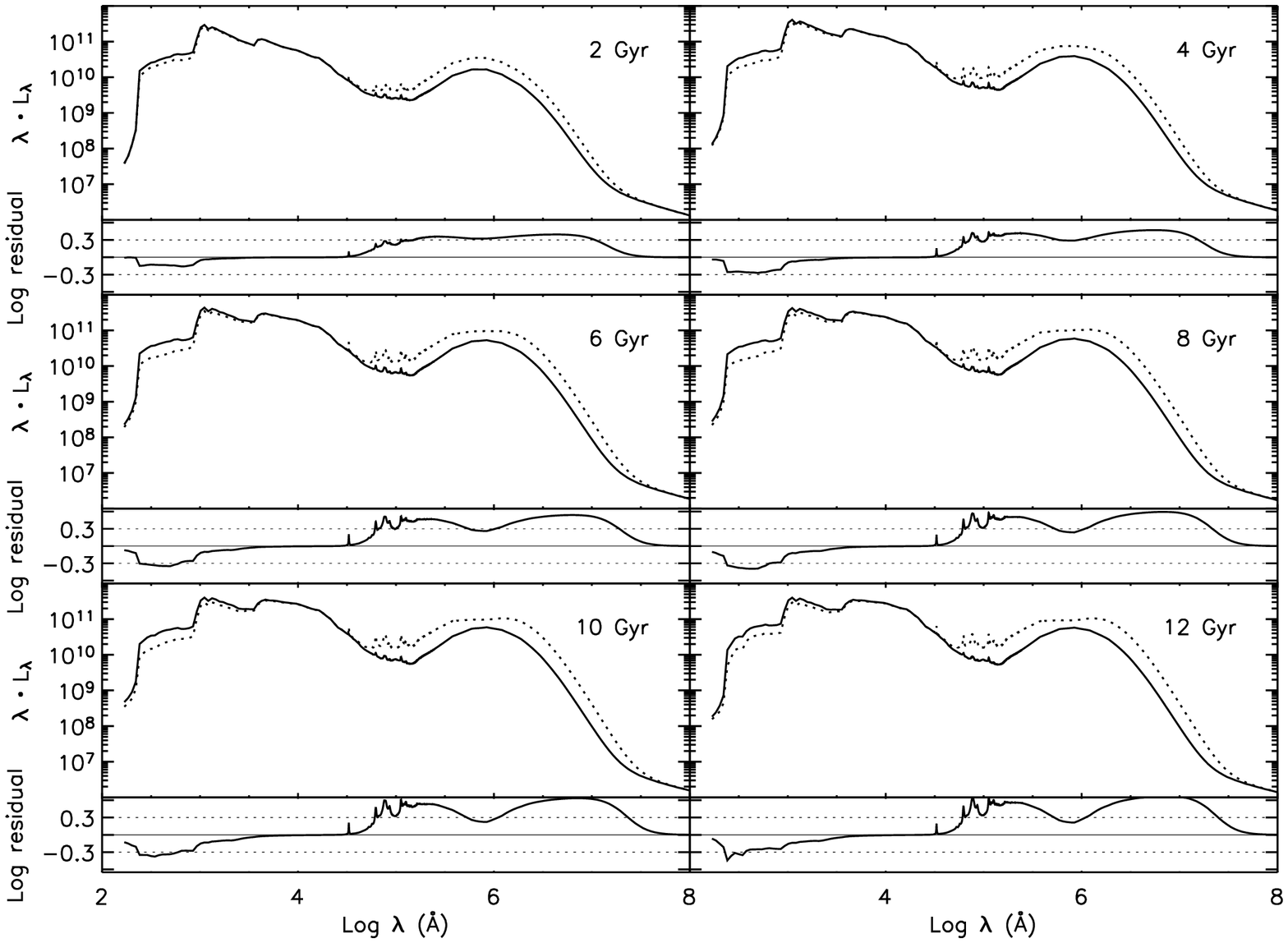}\\
 \caption{The evolution of the SED of the irregular galaxy, panels the same as for figure~\ref{g}}
\label{l}
\end{figure*}

\begin{figure}
%%  % Requires \usepackage{graphicx}
 \includegraphics[type=eps,ext=.eps,read=.eps, width=6cm,angle=90]{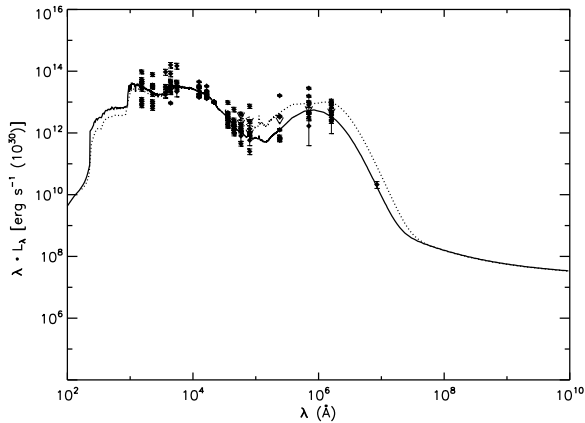}\\
 \caption{Comparison of theoretical irregular model at 12 Gyrs with SEDs of all irregular galaxies of magellanic type in the SINGS sample.
Solid line - Full CPM08 model. Dotted line - simple dust model}
\label{j}
\end{figure}

Figure~\ref{m}a shows that at ages greater than about 2 Gyrs the
chemical composition calculated by the CPM08 model is significantly
different from that calculated for the solar neighbourhood,
resulting in a much lower C/Si ratio. Figure~\ref{k}b shows that the
main effect on the SED of the galaxy of this lower C/Si ratio will
be in the MIR part of the spectra, with only a small effect at other
wavelengths. The effect on the MIR of a change in the chemical
composition needs to be treated with caution, since a major
contributor to this part of the spectra are PAHs, whose evolution is
not specifically followed by the CPM08 models (see section
\emph{4.2}). The effect of the lower C/Si value predicted by CPM08
on the SED will largely be due to a reduction in the number of PAH
molecules and therefore in the luminosity radiated in the PAH bands.

Although the treatment of the PAHs in this work is insufficient to
put any strong constraint on the nature of the PAHs in the galaxies
modeled, it is worth noting the relatively low abundance of C/Si
predicted for the irregular galaxy, compared to that of the solar
neighbourhood. This is interesting since it is in agreement with the
striking observed correlation between the strength of PAH features
and the metallicity of the galaxy (see e.g. Engelbracht et al. 2005;
Madden et al. 2006; Wu et al. 2006; O'Halloran et al. 2006), with
low metallicity galaxies, such as our theoretically modeled
irregular galaxy, exhibiting very weak or no features. Several
different explanations have been proposed for this correlation.
These include the suggestion of more efficient PAH destructive
mechanisms in lower metallicity galaxies (e.g. Galliano 2005; Madden
et al. 2006) due to harder radiation and the paucity of dust.
Alternatively, O'Halloran et al. (2006) proposed that the PAHs could
be destroyed by numerous shocks observed in low metallicity systems.
While it has also been shown by Galliano et al. (2008), that it is
possible to explain the trend of PAH abundance with metallicity,
using chemical evolution models. Starting from the summations that
lower metallicity galaxies are on average younger than higher
metallicity galaxies and that carbon dust and PAHs are predominantly
produced in AGB stars which recycle their ejecta into the ISM after
a relatively long period of time. It follows that the trend can be
seen as a reflection of this delayed injection of carbon dust into
the ISM by AGB stars, since younger systems will have had
insufficient time to become enriched with PAHs. The CPM08 chemical
evolution models used in this work puts forward a new suggestion
since they show that even in older low metallicity systems there is
likely to be a lower abundance of carbon and therefore PAH
molecules, a possibility which could explain or at least contribute
to the observed trend.

The relatively lower value of C/Si predicted by the CPM08 model with
respect to that predicted for the solar neighbourhood can be
explained by examining the different production and destruction
rates used in the two different environments. In the solar
neighbourhood model, carbon is assumed to be destroyed with a higher
efficiency than the silicates. This was chosen in order to explain
the local dust abundance pattern (see CPM08), and is consistent with
the fact that carbon has a condensation temperature lower than many
silicates. However, the solar neighbourhood model also includes dust
accretion, whose rate is assumed to be independent of the particular
element. For this reason, the abundance pattern produced by the
preferential destruction of certain elements is partially smoothed
out by the non-differential accretion. In the irregular model,
however, where dust accretion is absent, the higher destruction rate
attributed to the carbon is not compensated in the same way, giving
rise to the lower C/Si value.

A possible source of error in this work is the absence of an
additional 'very cold grain' (VCG) component as proposed by Galliano
et al. (2003 and 2004). In these papers a population of grains at
very low temperatures were introduced in order to explain an excess
in the emission in the sub-mm in the low metallicity galaxies
observed. Such a component could be located at the center of dense
clouds (see e.g. Galliano et al. 2003) or in the galaxies outskirts
where the radiation field is weak (see e.g. Draine et al. 2007).
Galliano et al. (2003 and 2005) estimated that for four low
metallicity irregular galaxies between 40 and 80 per cent of the
total dust mass is in the form of VCGs. Such a dust component is not
currently included in our models and if confirmed by additional
sub-mm observations the inclusion of such a large cold dust grain
population should be investigated. However any effect is likely to
be slight. For the irregular galaxies over 80 per cent of the
reprocessing is taking place within the MCs, heated by the young
stars (As opposed to about 25 per cent in the spiral model). The
emission of which is characterised only by the geometry and
dust-to-gas ratio of the MCs and would therefore be unaffected by a
population of VSGs. If a substantial proportion of the rest of the
dust takes the form of VCGs the result would be a reduction in the
amount of other larger sizes of dust grain in the cirrus. In our
irregular model the cirrus component only has a small observed
effect on the SED at wavelengths longer than the peak IR wavelength.
Therefore any effect will be slight and will probably result in a
slight decrease in emission in the very FIR and an increase in the
sub-mm, where the VSGs will emit.

\begin{figure}
%%  % Requires \usepackage{graphicx
 \includegraphics[type=eps,ext=.eps,read=.eps, width=6cm,angle=90]{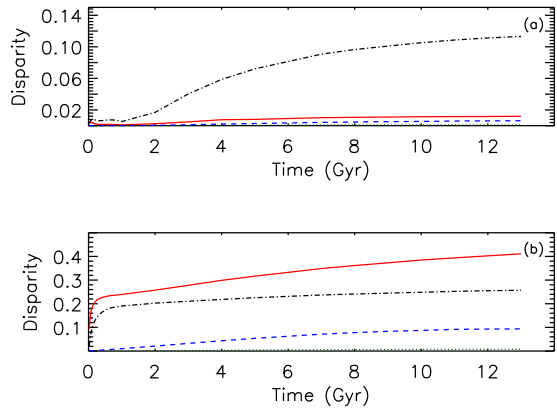}\\
 \caption{The residuals between the SEDs generated by the full irregular CPM08 model and SEDs generated by using the simplifications that
(a) the chemical composition remains equal to the Milky Way values
at all times, (b)the dust-to-gas ratio scales with the metallicity.
Lines the same as in figure 9} \label{k}
\end{figure}

The evolution of the SED of irregular galaxies is shown in
figure~\ref{l}. The quiescent low star formation rate leads to SEDs
which only slightly evolve with time. The mass of dust present in
the galaxy initially increases rapidly until a time $\sim$ 1 Gyrs is
reached at which point it begins to plateau (see figure~\ref{c}c).
The fraction of reprocessed light in the galaxy follows suit, rising
rapidly to a value of $\sim$ 15 per cent and then remains fairly
constant throughout its lifetime. A value which is one half of the
predicted light reprocessed in spiral galaxies.

As figure~\ref{k} shows the disparity between the SEDs generated
using the simplifications and the full dust treatment actually
decrease gradually as the galaxy gets younger with the difference
from introducing the dust-to-gas scaling relation always dominating
that from assuming a constant chemical composition. With the largest
effect always in the IR with a very small effect in the optical and
NIR part of the spectrum.

\begin{figure}
  \centering
  \includegraphics[type=eps,ext=.eps,read=.eps, height=8.5cm, angle=90]{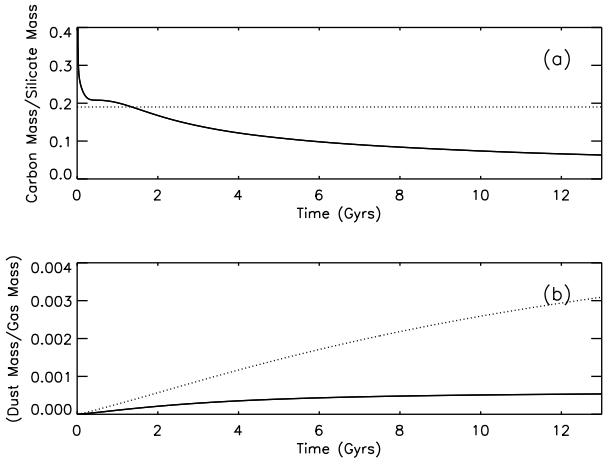}\\
  \caption{Irregular Galaxy: (a) Comparison between the value of the C/Si calculated by the full CPM08 ratio (solid line) with
the constant value assumed if the chemical abundance remains
identical to the present value for the solar neighbourhood (dotted
line). (b) Comparison of the dust-to-gas ratio value calculated by
the full CPM08 model (solid line) with the expected value if a
dependence on the metallicity is assumed (dotted line).} \label{m}
\end{figure}
\begin{figure}
%%  % Requires \usepackage{graphicx}
 \includegraphics[type=eps,ext=.eps,read=.eps, width=6cm,angle=90]{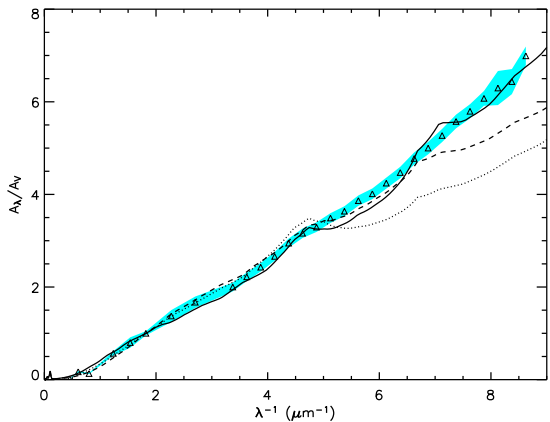}\\
 \caption{Best fit dust composition to SMC extinction curve. Solid line - full CPM08 model with optimised size
distribution; dashed line - full CPM08 with S98 size distribution;
dotted line - Milky Way chemical composition, S98 size distribution
. The blue shaded area shows the associated uncertainty in the
observed extinction curve.} \label{n}
\end{figure}
The dust size distribution in irregular galaxies is likely to differ
from that of the Milky Way, (e.g. Galliano et al. 2005). This is
highlighted by the fact that extinction curves have been observed in
both the small magellanic cloud (SMC) and large magellanic cloud and
have been found to differ from that of the Milky Way. The SMC
extinction curve is of particular interest since it is an irregular
dwarf galaxy with a substantially lower metallicity than the Milky
Way and should represent a dust environment similar to our
theoretical irregular model. The predicted extinction curve for the
mass and composition of dust predicted by our irregular model has
been compared to the observed SMC curve in order to constrain a dust
grain size distribution. As figure~\ref{n} shows just by adopting
the chemical composition predicted by the full CPM08 model even with
the size distribution already calculated to match the properties of
the Milky Way a reasonable fit is obtained to the extinction curve,
so the different chemical composition alone can go some way to
explaining the observed extinction curve. By fine tuning the
parameters governing the size distribution in equation 7 an improved
fit can be obtained, as shown in the figure. The values of the best
fit parameters are shown in table 1.

\begin{figure*}
 \includegraphics[type=eps,ext=.eps,read=.eps, width=13cm,angle=90]{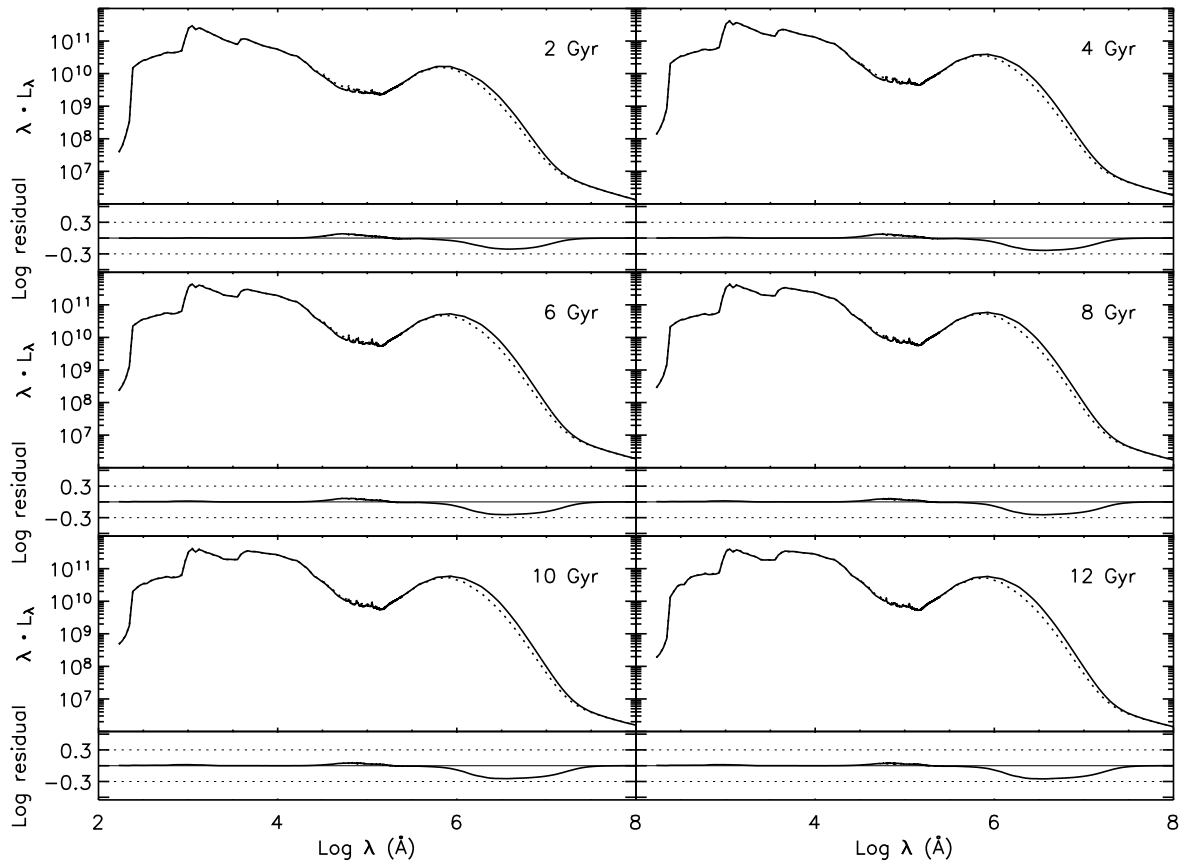}\\
 \caption{The effect of the SMC size distribution on evolution of the SED of the irregular galaxy, solid line full CPM08 model with MW size
distribution, dashed line - full CPM08 model with SMC size
distribution} \label{o}
\end{figure*}

Figure~\ref{o} displays a comparison of the SEDs for the full model
adopting the size distribution constrained using the SMC extinction
curve to the SEDs for the full model using the standard Milky Way
derived size distribution. It shows that the difference is small at
all times, largely because the difference in the two size
distributions is small.

\subsection{Properties of Elliptical Galaxies}
In our elliptical galaxy formation scenario there are two distinct
stages. An initial rapid collapse of a sphere of primordial gas
leading to intense star formation and a second quiescent stage where
the star formation has stopped as almost all of the gas is blown
away by galactic winds. These two stages will be discussed
separately. Unless stated all results shown are for the elliptical
model with a final stellar mass of $10^{11} M_{\sun}$.
\subsubsection{Starbursting phase - High redshift}

\begin{figure*}
 \includegraphics[type=eps,ext=.eps,read=.eps, width=13cm,angle=90]{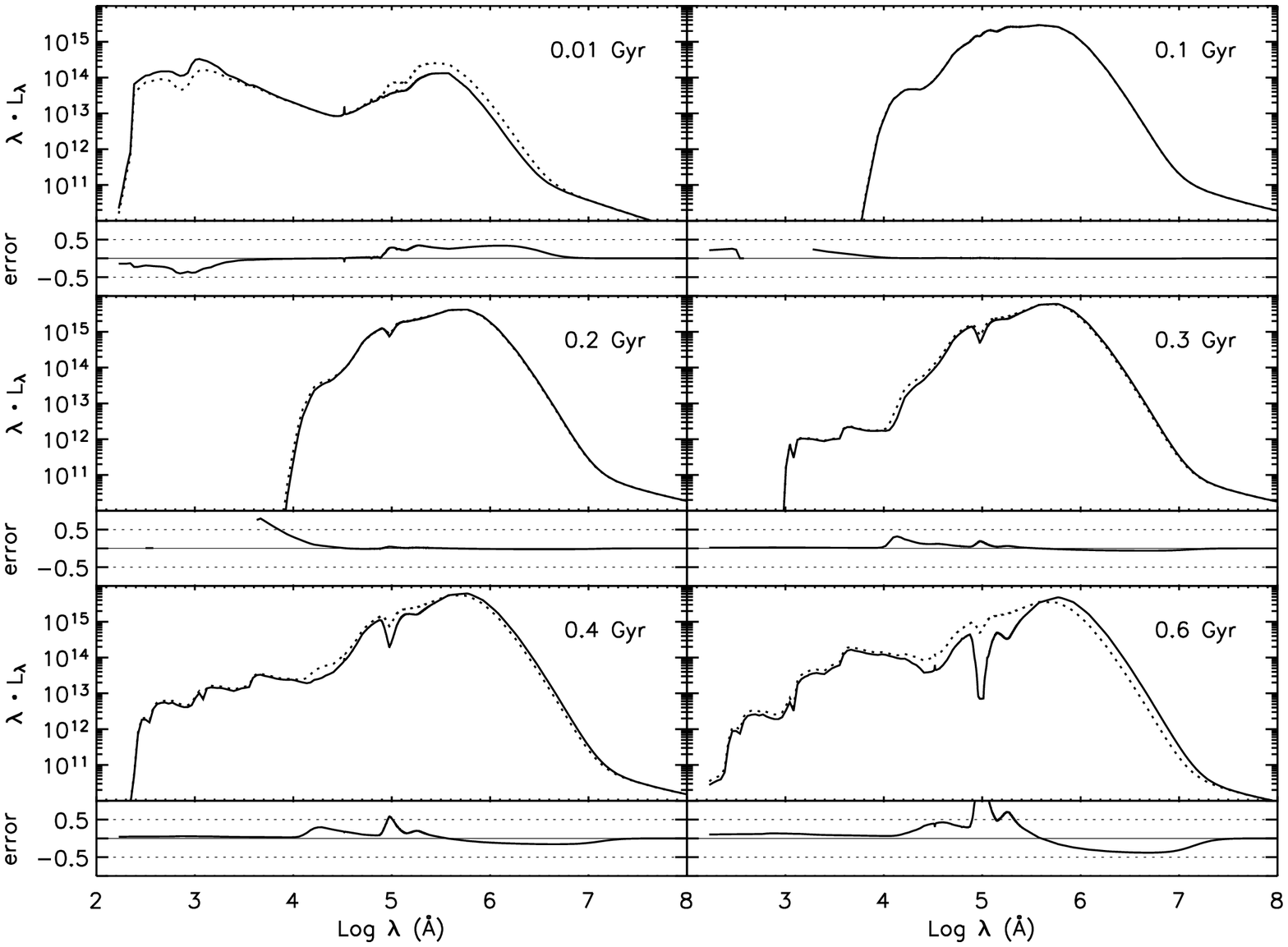}\\
 \caption{The high redshift evolution of the SED of the elliptical galaxy, panels the same as for figure~\ref{g}}
\label{t}
\end{figure*}

The evolution of the SED of our elliptical model during the
starbursting phase is shown in figure~\ref{t}. Initially all the
stars will be located within their birth clouds (MCs) however since
the dust mass is initially negligible the SED at these very early
times will have the form of just the bare stellar component. The MCs
will soon become rapidly enriched with dust due to dust production
in type II SNe and as a result the reprocessing will rise quickly,
almost to 100 per cent after 0.005 Gyrs (figure~\ref{f}). As the age
of the galaxies increase more stars will have had time to escape
from their MCs and as a result will be radiating in the more diffuse
cirrus hence the gradual decrease in reprocessing as observed in the
figure. The small amount of PAH emission seen in the SEDs of
figure~\ref{t} is as a result of the large proportion of dust
emission coming from the MCs. Due to the suppression of the
abundance of PAHs in these environments they will consequentially
produce little PAH emission. After the onset of the galactic winds
at 0.7 Gyr the majority of the dust is blown out of the galaxy so
the amount of reprocessing drops, at same the time the star
formation is stopped and the galaxy enters the passively evolving
stage.

\begin{figure}
 \includegraphics[type=eps,ext=.eps,read=.eps, width=6cm,angle=90]{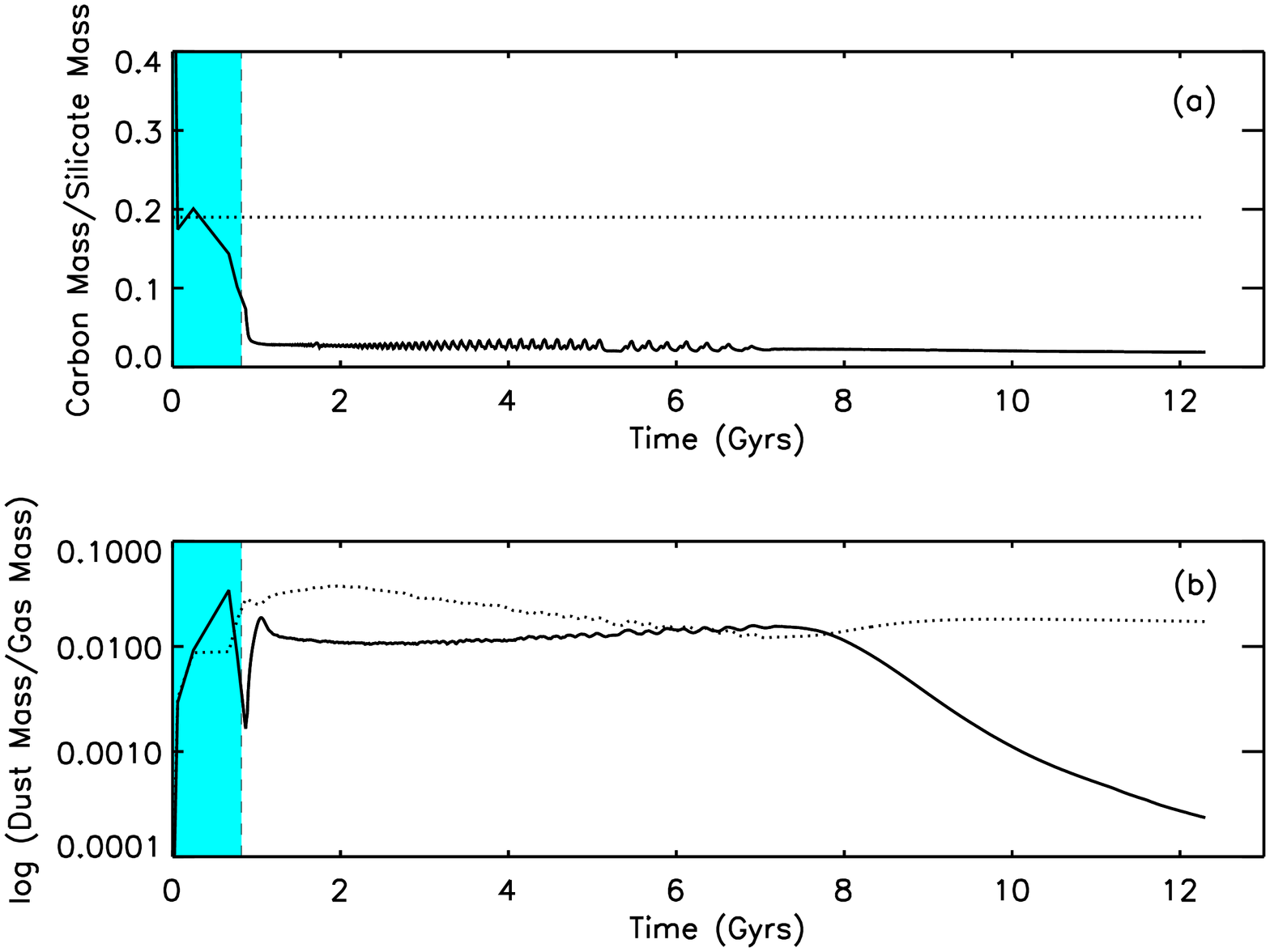}\\
 \caption{Elliptical Galaxy: (a) Comparison between the value of the C/Si calculated by the full CPM08 ratio (solid line) with
the constant value assumed if the chemical abundance remains
identical to the present value for the solar neighbourhood (dotted
line). (b) Comparison of the dust-to-gas ratio value calculated by
the full CPM08 model (solid line) with the expected value if a
dependence on the metallicity is assumed (dotted line). The blue
shaded area represents the age before the galactic wind starts in
the elliptical model during which it is undergoing a period of
extreme starformation} \label{cc}
\end{figure}
\begin{figure}
%%  % Requires \usepackage{graphicx
 \includegraphics[type=eps,ext=.eps,read=.eps, width=6cm,angle=90]{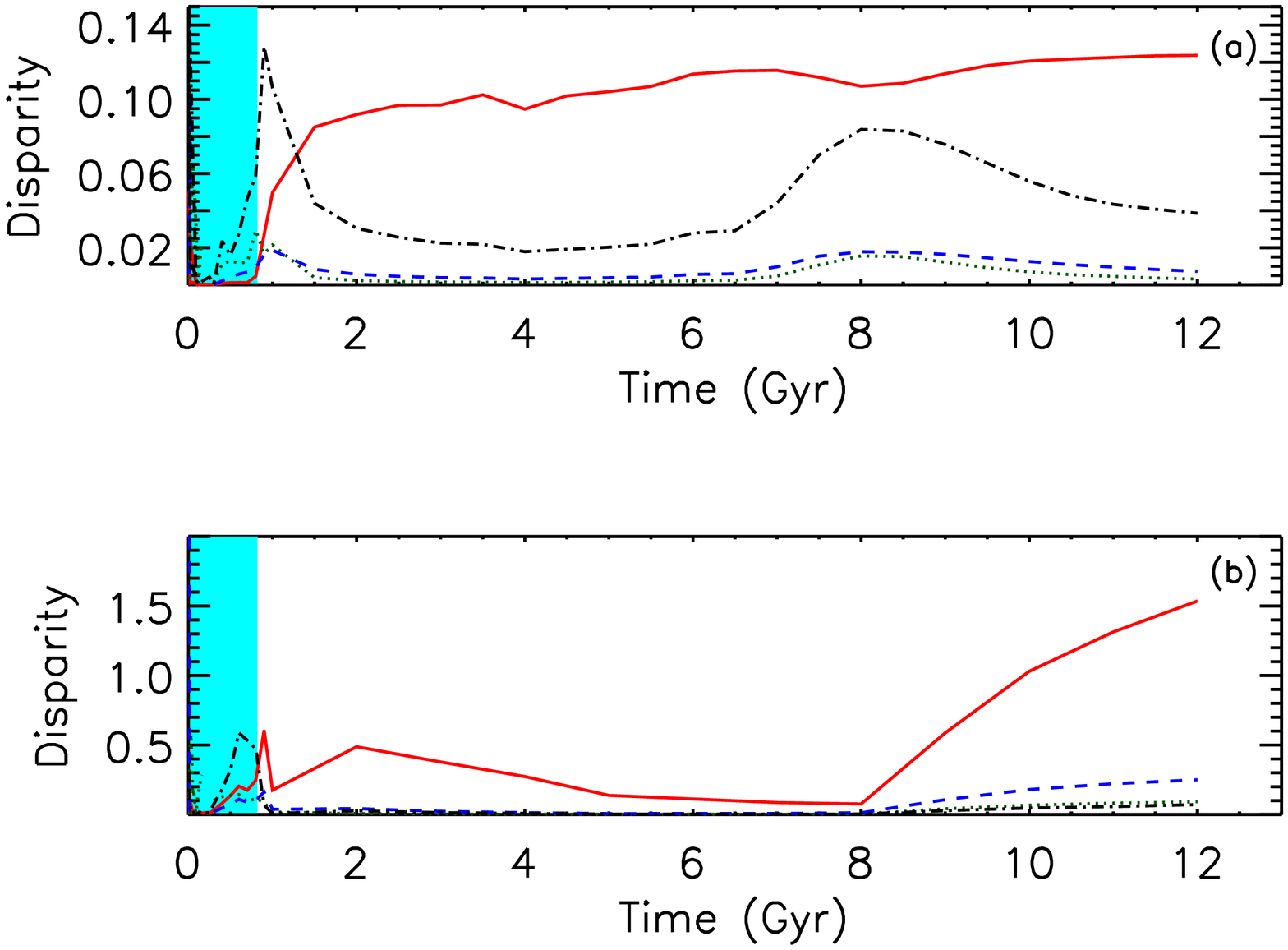}\\
 \caption{The residuals between the SEDs generated by the full elliptical CPM08 model and SEDs generated by using the simplifications that
(a) the chemical composition remains equal to the Milky Way values
at all times, (b)the dust-to-gas ratio scales with the metallicity.
Lines the same as in figure 9. The blue shaded area represents the
age before the galactic wind starts in the elliptical model during
which it is undergoing a period of extreme starformation} \label{r}
\end{figure}

Figure~\ref{r} shows the effect that the two simplifications will
have on the SEDs. Of the two, adopting the assumption that the
dust-to-gas ratio will scale with the metallicity will have the
biggest effect. This is because as figure~\ref{cc} shows that
although such a relationship does hold for these young galaxies
initially (ages $<$ 0.3 Gyrs) the dust-to-gas value calculated by
the model soon diverges from the expected value based on the
metallicity and as a result using this simplifications leads to
large disparities in the SEDs.

Assuming a milky way chemical composition for the dust however has
little effect on the SEDs at early times. This is because as
figure~\ref{cc} shows the chemical composition at least initially is
remarkably similar to the value of 0.19 calculated for the solar
neighbourhood. As the burst progresses the $C/Si$ begins to drop
leading to small differences between the SEDs of the full CPM08
model and the simple dust model (see figure~\ref{r}b).

Although the model calculates that the chemical composition of the
total dust component is likely to be quite similar to that of the
solar neighbourhood throughout the burst it does not necessarily
follow that the individual grains will be of the same composition
and have the same size distributions as that observed locally. In
the local universe much of the dust is thought to be produced in the
stellar winds of low and intermediate mass stars. However in young
galaxies at high redshift with ages less than 1 Gyr such stars would
not have had sufficient time to evolve to this stage, instead the
dust will be predominantly produced in SNe, for detailed information
of the relative production rates of the various mechanisms see
figures 2 and 6 in CPM08. Hence the dust properties in these young
galaxies is thought to be very different to that observed in the
local universe in the Milky Way and in the SMC.

One possible tool for the study of dust in the high redshift
universe is QSOs. According to recent models of QSO-galaxy
co-evolution (e.g. Granato et al. 2004 and Di Matteo et al. 2005)
during the early stages of galaxy and black hole growth the AGN is
embedded in gas and dust and is therefore difficult to identify as
an QSO, however after the onset of galactic winds because large
quantities of dust and gas will be swept away the QSO can be
observed. The properties of observed QSOs should therefore be
predicted by our elliptical model at times shortly after the onset
of the galactic wind. Maiolino et al. (2006) showed that the
chemical abundances observed in QSOs best match the chemical
abundances calculated for our giant $10^{12} M_{\sun}$ elliptical
model at times later than 0.5 Gyrs corresponding to the epoch just
after the onset of the galactic winds supporting this theory. By
studying the SEDs of QSOs it has been possible to infer extinction
curves for these objects and by studying these observed curves it
should be possible to constrain a size distribution for our young
elliptical galaxies.

Maiolino et al. (2004) calculated an extinction curve for a high
redshift QSO, SDSSJ1048+46 at a redshift of z=6.193 and found that
the extinction curve  differs greatly from the extinction curves
observed at low redshift. The observed extinction curve has been
compared to the extinction curve predicted by several models which
have been written to predict the size distribution of dust from SNe
including, Bianchi \& Schneider (2007) and Nozawa et al. (2007).
These size distributions have been found to be consistent with the
observered QSO extinction curve, leading to the conclusion that the
dust observed in the extremely high redshift objects could indeed be
dominated by SNe dust. The processes involved in the formation of
dust in SNe is complicated and is far from being completely
understood. Indeed the size distributions predicted by Nozawa et al.
(2007) and Bianchi \& Schneider (2007) differ substantially from
each other. In addition both papers only consider the production and
destruction of dust in SNe, neither follow the accretion and
destruction of the SNe remnants in the ISM as well as that by the
galactic winds.

\begin{figure}
%%  % Requires \usepackage{graphicx}
 \includegraphics[type=eps,ext=.eps,read=.eps, width=6cm,angle=90]{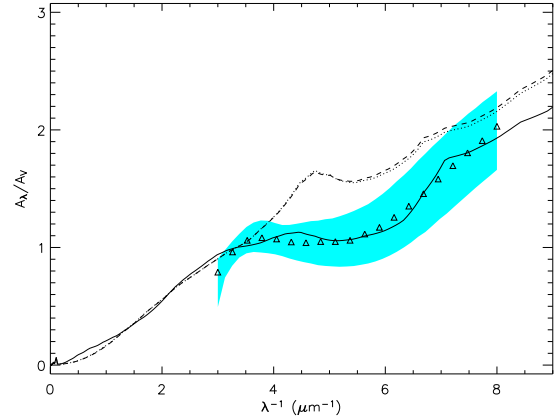}\\
 \caption{Best fit dust composition to the SDSS 1048
+46 quasar extinction curve. Solid line - full CPM08 model with
optimised size distribution; dashed line - full CPM08 with S98 size
distribution; dotted line - Milky Way chemical composition, S98 size
distribution. The blue shaded area shows the associated uncertainty
in the observed extinction curve.} \label{w}
\end{figure}

In this paper starting from the dust masses calculated by the CPM08
$10^{12} M_{\sun}$ elliptical model, at a galactic age of 0.5 Gyr,
the parameters in equation 7 which correspond to the best fit to the
Maiolino et al. (2004) extinction curve at Z$\sim$6 have been
calculated. The resultant extinction curve is shown in
(figure~\ref{w}). The best fit parameters are shown in table 1.

\begin{figure*}
 \includegraphics[type=eps,ext=.eps,read=.eps, width=13cm,angle=90]{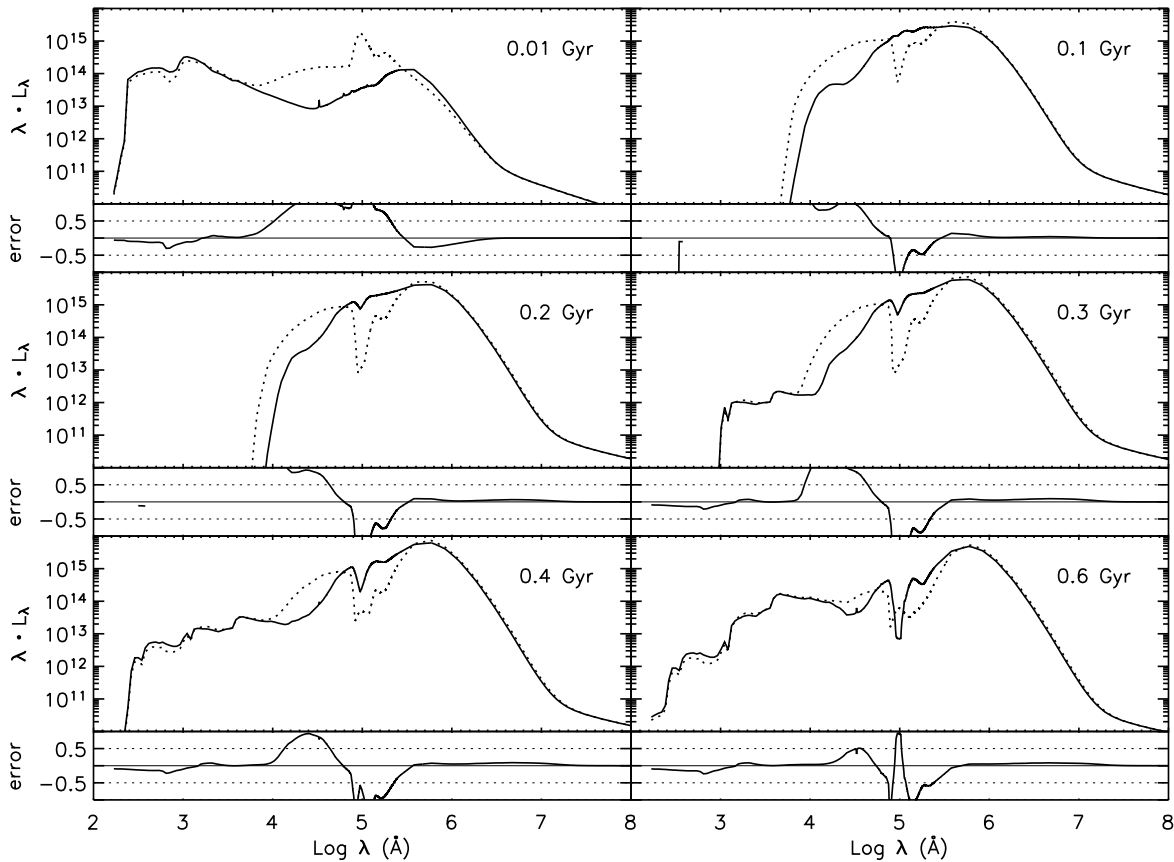}\\
 \caption{The effect of the quasar size distribution on evolution of the SED of the young elliptical galaxy, solid line full CPM08 model with MW size
distribution, dashed line - full CPM08 model with the dust
consistent with the z$\sim6$ QSO.} \label{x}
\end{figure*}

Figure~\ref{x} shows the effect that adopting this size distribution
will have on the SEDs of elliptical galaxies during their starburst
phase. The SEDs generated vary substantially from those generated
using the Milky Way size distribution particularly in the NIR and
MIR.\\

\noindent \emph{{SCUBA galaxies}}

The theoretical SEDs for the $10^{11} M_{\sun}$ young elliptical
galaxies are compared to SCUBA galaxies at high redshifts. These
galaxies have been found to be dust enshrouded sources with very
high star formation rates. It is believed that these SCUBA galaxies
could represent the birth of giant elliptical galaxies (Lilly et al.
1999, Eales et al. 2000) and should therefore correspond to the
starbursting phase of our elliptical models. The data used is from
the SHADES survey of the SXDF field (Clements et al. 2008). This
survey is ideal for the purposes of this paper since it has a good
photometric coverage with observations from the optical Suburu /
XMM-Newton field Deep Field (SXDF) (\emph{B}, \emph{V}, \emph{R}
\emph{i'} and \emph{z'}), Spitzer (IRAC 3.6, 4.5, 5.8, 9 $\mu$m and
MIPs 24, 70 and 160 $\mu$m) and SCUBA. Only the galaxies with the
highest redshift from the sample were used, namely galaxies which
were found by Clements et al. (2008) to have a photometric redshift
greater than two. In total these galaxies numbered thirteen.

Figure~\ref{cc} shows that for four of the observed galaxies in the
SHADES dataset our model was able to match the general shape of the
SED and the observed amount of reprocessing. The average age of the
best fitting full CPM08 models for the four galaxies shown is 0.55
Gyrs. At such an age the theoretical model will still have ongoing
star formation. Because only one star formation history has been
followed and no parameters varied perfect matches could not be
expected to all observed galaxies. However, as shown, even with this
one simple star formation history, the observed SEDs of four out of
the thirteen galaxies investigated in this work could be reasonably
reproduced. This suggests that the rationale of treating these
galaxies as obscured starbursts with a dust mass calculated by the
CPM08 model is consistent with the observations. Also shown in the
figure are the spectra for the simple dust model and for the full
dust model with the dust composition calculated in order to fit the
QSO SDSSJ1048+46. Although the SEDs generated using the full model
with the size distribution consistent with the z$\sim$6 QSO diverge
from that of the full CPM08 model significantly, the theoretical
SEDs are still largely consistent with the data. The SEDs generated
by the simple dust model on the other hand lead to slightly worse
fits, although a much more thorough analysis is needed to come to
a more definite conclusion.\\

\noindent \emph{Giant post-starbusrt galaxies at high redshift}

The theoretical SEDs calculated for young elliptical galaxies are
compared to massive galaxies which have been observed to be already
highly evolved at high redshifts (Wiklind et al. 2008). Since these
galaxies are likely to be massive the giant $10^{12} M_{\sun}$
elliptical galaxy will be used in the fits. The galaxies in the data
set have been selected by Wiklind et al. (2008), so that they are
dominated by a stellar population older than $\sim$ 100 Myr and
situated at z $\geq$ 5. They should therefore correspond to a later
stage in the evolution of our elliptical galaxy model, the post
starburst stage of our giant elliptical model. Data points are
available in the optical (ACS - \emph{B}, \emph{V}, \emph{i},
\emph{z}) NIR (VLT/ISAAC - J H $K_{s}$) - MIR, deep imaging with the
Spitzer Space Telescope with IRAC (3.6, 4.5, 5.7 and 8.0 $\mu$m) and
MIPS (24 $\mu$m). For the chi-squared fitting the point at 24 $\mu$m
was not included since emission in this region could be dominated by
an AGN.

\begin{figure*}
%  % Requires \usepackage{graphicx}
\includegraphics[type=eps,ext=.eps,read=.eps, width=13cm,angle=90]{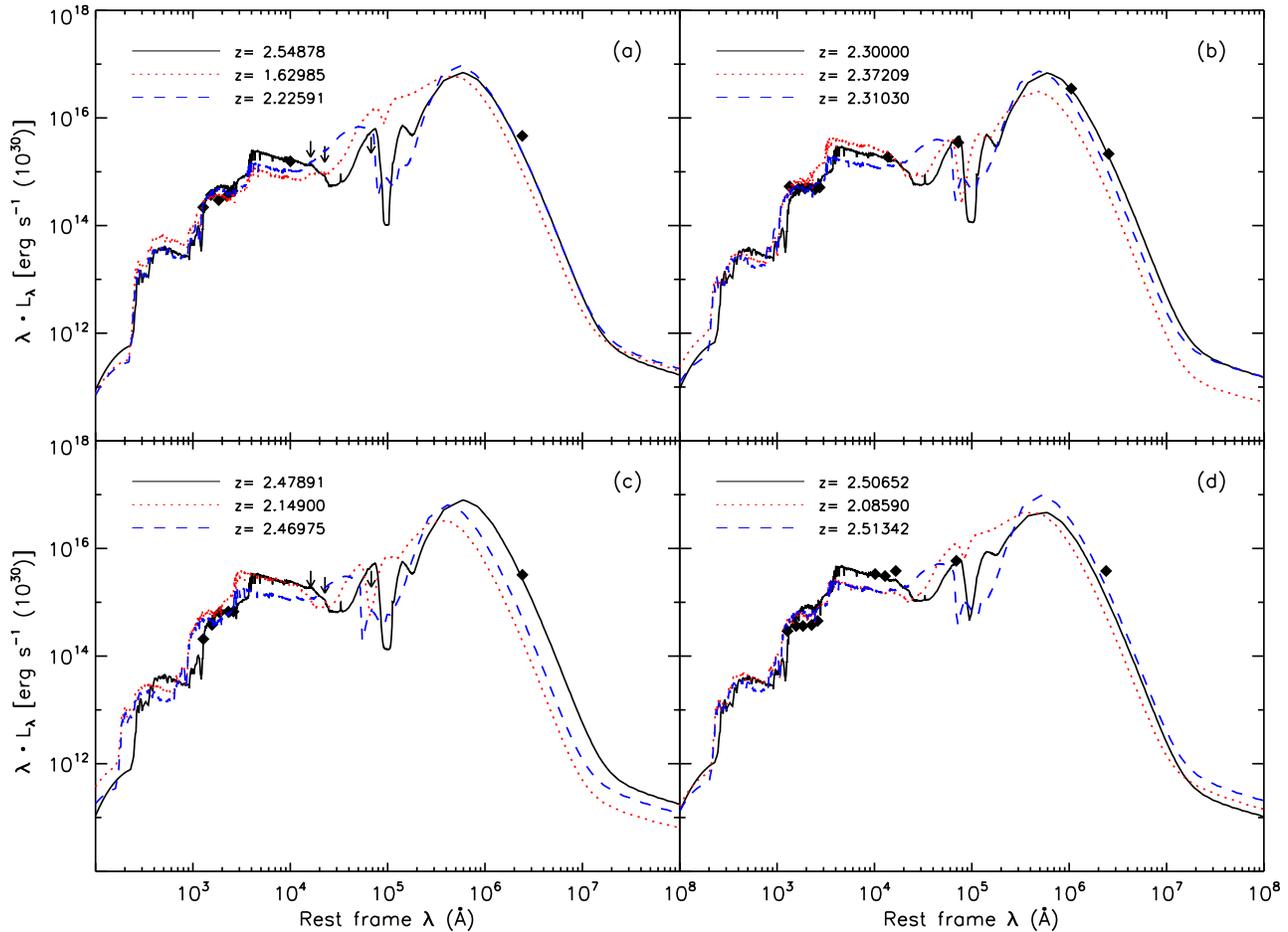}\\
 \caption{Comparison of $10^{12} M_{\sun}$ theoretical elliptical model with the SED of SCUBA galaxies: a) SHADES-SXDF 10, b) SHADES-SXDF 11, c) SHADES-SXDF 12, d) SHADES-SXDF 7.  Solid line - full CPM08 model. Dotted line - simple dust model. Dashed line - full CPM08 model with dust consistent with the z$\sim$6 QSO.
 The rest frame wavelength is correct for the redshift calculated for the full CPM08 model. The abscissa values for the curves for the simple dust model and the full CPM08 model with dust consistent with the z$\sim$6 QSO were appropriately re-scaled. The redshift values calculated by Clements et al. (2008)
 are:  z= 2.08 for SHADES-SXDF 10, z=2.30  for SHADES-SXDF 11, z=3.07 SHADES-SXDF 12, z=2.31 SHADES-SXDF 7}
\label{y}
\end{figure*}

\begin{figure*}
%  % Requires \usepackage{graphicx}
\includegraphics[type=eps,ext=.eps,read=.eps, width=13cm,angle=90]{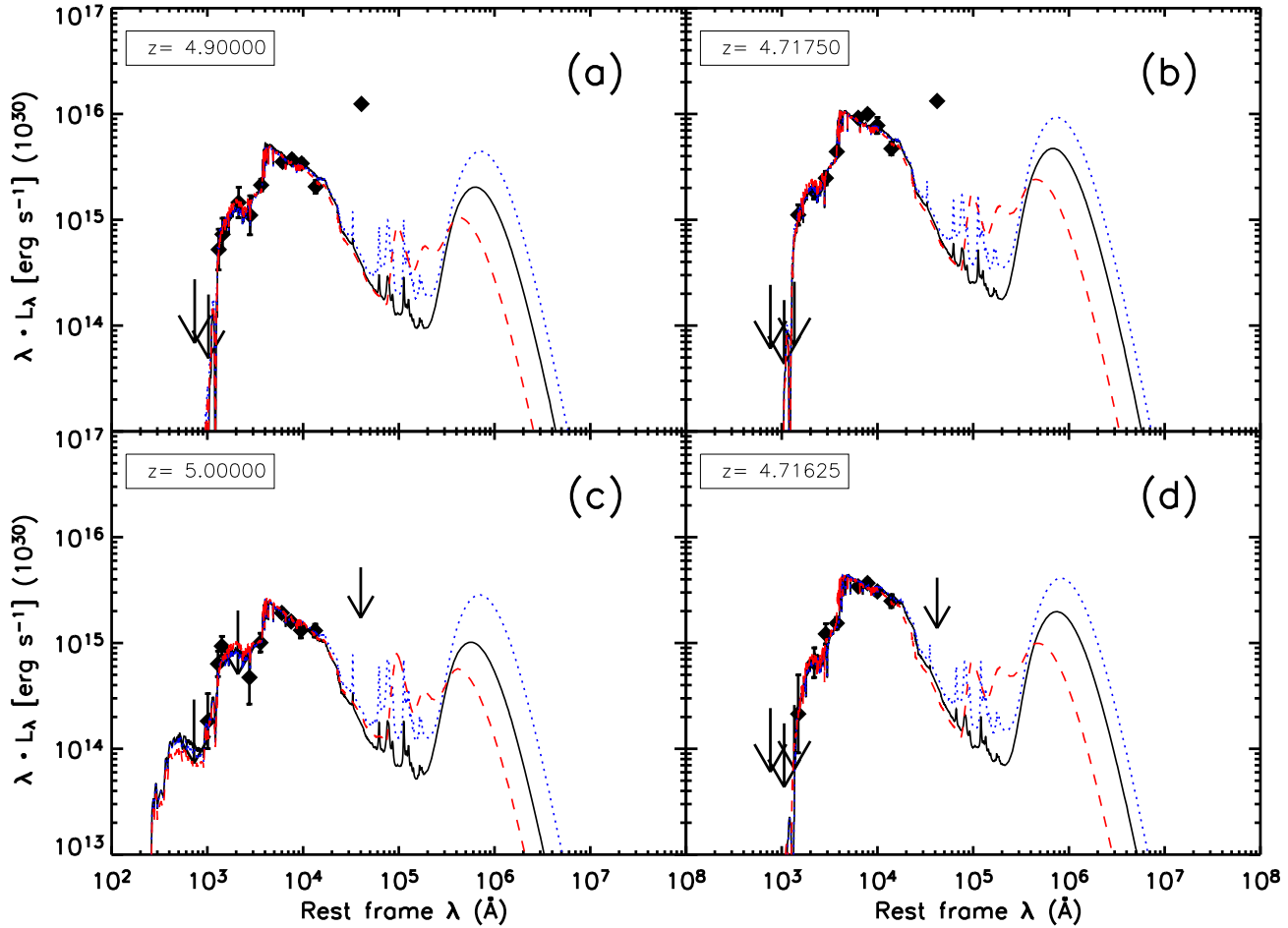}\\
 \caption{Comparison of the $10^{12} M_{\sun}$ theoretical elliptical model with the SED of balmer break galaxies: a) BBG 2910, b) BBG 3348, c) BBG 3361, d) BBG 4071.  Black solid line - full CPM08 model. Blue dotted line - simple dust model. Red dashed line - full CPM08 model with dust consistent with the z$\sim$6 QSO. The rest frame wavelength
 is correct for the redshift calculated for the full CPM08 model, the value given in the top left corner of the panel. In all cases the redshift values were the same for the two other models,
 except for BBG 4071, where a redshift of 4.34875 was calculated for the simple dust model, the abscissa values for the curve were appropriately re-scaled. The redshift values calculated by Wiklind et al.(2008) are: z=4.9 for BBG 2910, z=5.1 for BBG 3348, z=5.0 for BBG 3361 and z=5.0 for BBG 4071.} \label{y}
 \label{cc}
\end{figure*}

\begin{figure}
\includegraphics[type=eps,ext=.eps,read=.eps, width=6cm,angle=90]{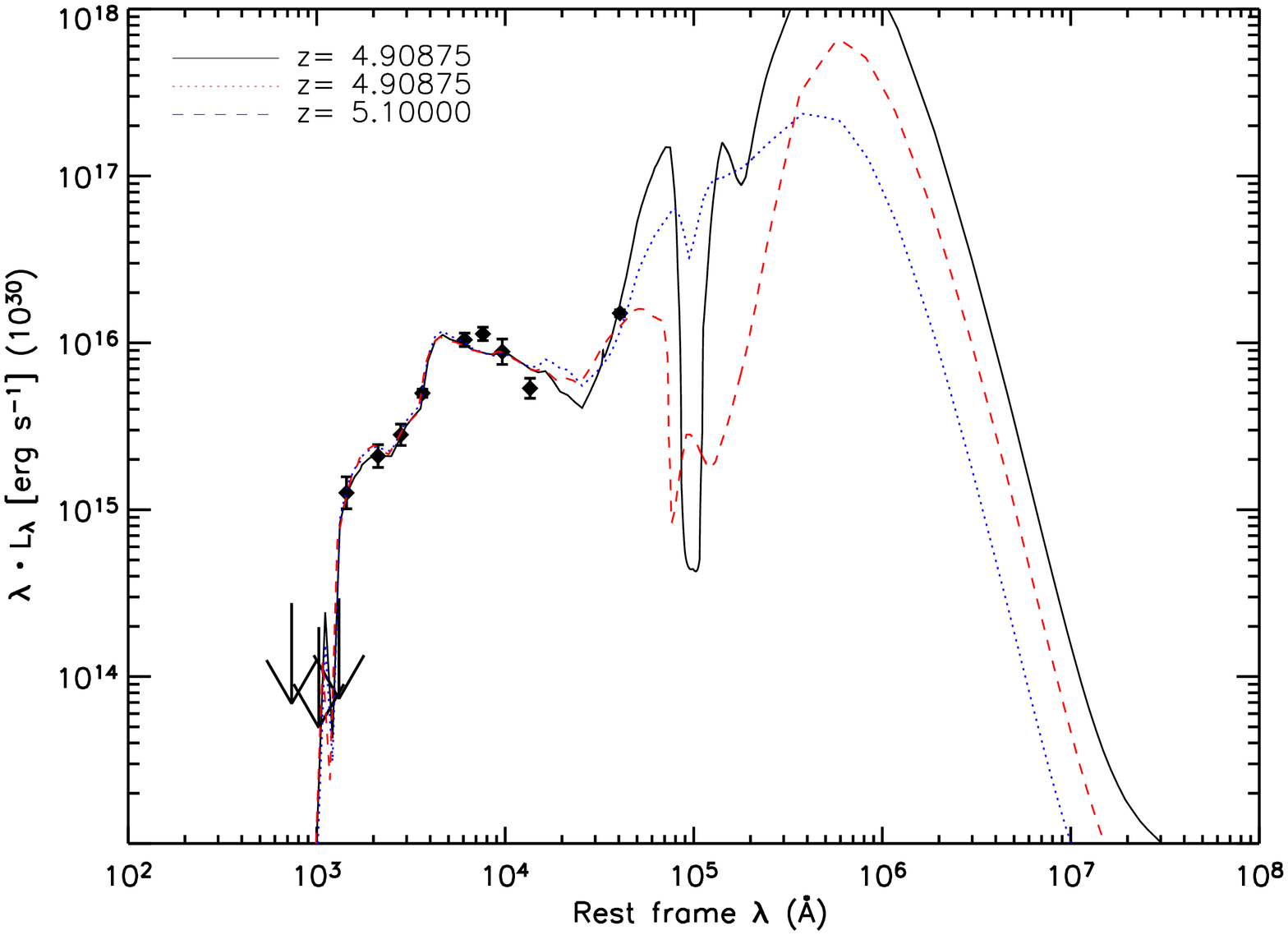}\\
 \caption{Comparison of the $10^{11} M_{\sun}$ theoretical model to galaxy BBG 3348 from the Wiklund sample.  Dotted line - simple dust model. Dashed line - full CPM08 model with dust consistent with the z$\sim$6 QSO.
 The rest frame wavelength is correct for the redshift calculated for the full CPM08 model. The abscissa values for the curves for the simple dust model and the full CPM08 model with dust consistent with the z$\sim$6 QSO were appropriately re-scaled.
Wiklind et al. (2008) calculated a redshift value of
z=5.0}\label{bb}
\end{figure}

Many of the observed galaxies could be well matched by our
theoretical model and figure~\ref{bb}a shows good fits to four of
the galaxies in the sample. The fits for the full CPM08, the simple
dust model and the full dust model with the dust composition
calculated in order to fit the QSO SDSSJ1048+46 are all good. The
theoretical SEDs diverge from each other mostly in the rest frame
FIR and since no observations exist for this part of the spectrum
the difference in fits in the SEDs is slight. The average age of the
best fitting full CPM08 models for the four galaxies shown is 0.61
Gyrs. At such an age our giant elliptical model corresponds to a
post starbursting system with no current star formation. It should
be noted that because only one star-formation history was followed
in this work our calculated value for the stellar ages of the
galaxies should be taken as an indication of one possible formation
scenario not as an actual value.

For several of the galaxies in the sample, intriguingly the observed
MIPS flux at 24$\mu$m far exceeds the flux predicted by our model,
for example BBG 3348. This galaxy has been observed in the x-ray
with a luminosity of $3\times10^{43} erg$ $s^{-1}$ and it is
therefore plausible that the observed MIPS flux could have a
contribution from an obscured AGN. Such a source could explain the
higher flux observed in this band but would have a minimal impact on
the part of the SED covered by the ACS/ISAAC/IRAC bands (see e.g.
Mobasher 2005) and would therefore not effect the presented SED fit
in this region.

Figure~\ref{bb}, however, shows an alternative formation scenario
for this galaxy, in which the predicted SED of the $10^{11}
M_{\sun}$ model is normalised and compared to the observed SED. The
best fit model has been found to have a stellar age of 0.6 Gyr, the
same age as that calculated for the $10^{12} M_{\sun}$ model. As can
be seen from the fit, the MIPS 24$\mu$m flux is predicted by the
model without the need of an obscured AGN. In this $10^{11}
M_{\sun}$ model, star formation is still ongoing at such an age. Due
to the nature of the dust model used, this young star formation will
be entirely obscured within dense MCs so will only contribute to the
flux in the rest frame MIR and FIR. The older stars that have been
able to escape the MCs however will be only slightly absorbed due to
the less dense cirrus producing a similar SED in the optical and NIR
as the $10^{12} M_{\sun}$ model described above.

Both scenarios described, obscured starburst or a passive galaxy
with an AGN are equally consistent with the data.  Longer wavelength
observations would be required to discriminate between them.
\subsubsection{Quiescent stage- low redshift}
\begin{figure}
 \includegraphics[type=eps,ext=.eps,read=.eps, width=6cm,angle=90]{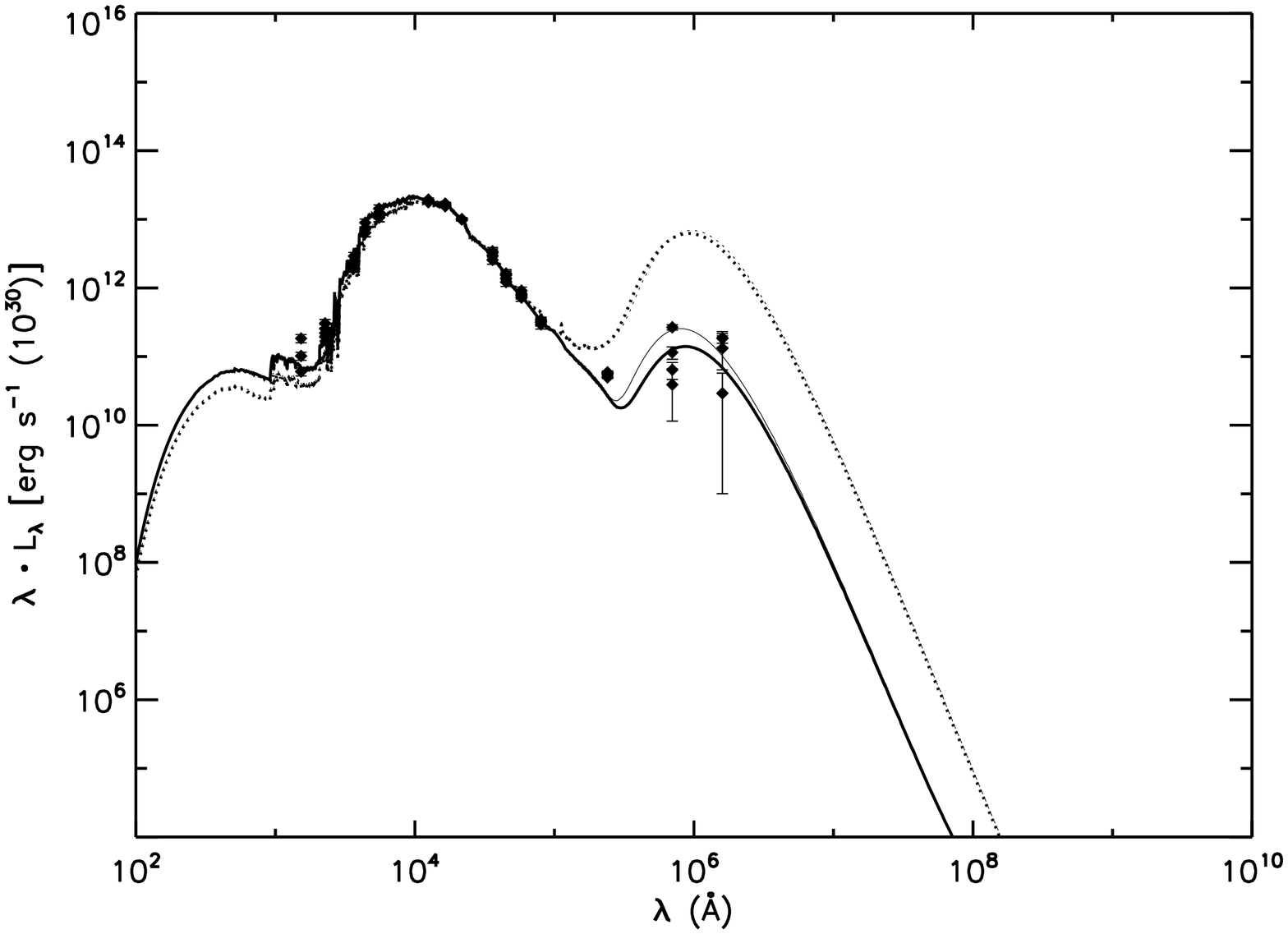}\\
 \caption{Comparison of theoretical elliptical model at 12 Gyrs with an average SED taken
from four elliptical galaxies of the SINGS sample. Solid line - Full
CPM08 model; dotted line - simple dust model. The thick line is for
a scalelength of dust twice as large as that of the stellar
component while the thin line is for equal scalengths. } \label{p}
\end{figure}

Figure~\ref{p} shows the result for the theoretical SED of the
elliptical model generated at 12 Gyrs compared to the SEDs of the
four 'passive' local elliptical galaxies from the SINGS sample. The
SED generated by the model with the full dust evolution matches the
observations well. Elliptical galaxies display fairly homogeneous
optical colours, a trend reflected in the observed galaxies in
figure~\ref{p}, which have almost identical SEDs from the optical to
the NIR. The SED in this region is reproduced well by our full
theoretical model and also the simple dust model. The four galaxies
though, show larger scatter in the FIR and in the UV.

Although there is much scatter in the observed FIR bump, probably
due to a differing dust content, the full CPM08 model predicts FIR
emission which agrees well with the average of the observed points.
It can therefore be thought of as having a dust content consistent
with this sample. However, when the simple dust model using the two
simple dust assumptions is used SEDs are generated which
significantly overestimate the amount of FIR observed. Presumably
because of the larger dust content present, due to the assumption
that the dust-to-gas ratio scales with the metallicity (see
figure~\ref{cc}b). In this work the assumption has been introduced
that the dust component is more extended than that of the stellar
component with a scalelength twice as large \emph{(see section
3.3.4)}. To check the robustness of our results on this assumption
the SED of the elliptical model at 12 Gyr with a scalelength of the
dust equal to that of the stellar component was calculated. This is
also shown in figure~\ref{p} and it can be seen that although this
assumption does have a noticeable effect on the SED the fit for the
full CPM08 model is still consistent with the observations while the
simple dust model stills over predicts the emission in the FIR.

Out of the four 'passive' elliptical galaxies in the sample all four
have observations in the NUV and three have observed points in the
FUV. Although the galaxies do show a small amount of scatter in this
region the theoretical model is able to reproduce a UV-upturn which
is consistent with the average points. The UV scatter too could be
partially explained by differing amounts of extinction due to
different dust contents. Indeed the simple dust model seems to
under-predict the luminosity in the FUV, presumably because of the
greater dust content present. However, it has also been claimed by
several authors (e.g. Kaviraj et al. 2007) that the range of UV
colours observed in different early type galaxies, could be due to a
small amount of recent star formation ($1\%-3\%$) in a large
proportion of galaxies. This however is beyond the scope of the
present paper.

\begin{figure*}
 \includegraphics[type=eps,ext=.eps,read=.eps, width=13cm,angle=90]{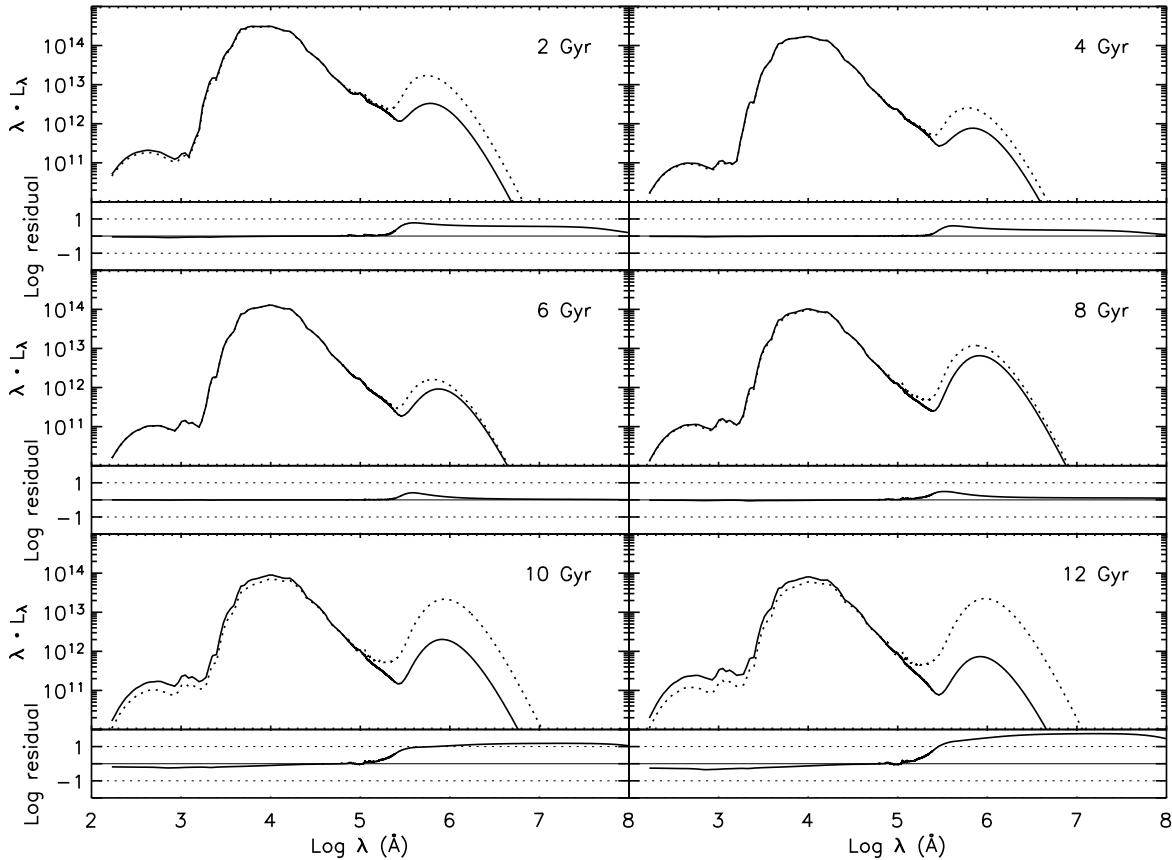}\\
 \caption{The evolution of the SED of the elliptical galaxy, panels the same as for figure 7}
\label{s}
\end{figure*}

The theoretical evolution of the SED of the elliptical galaxy is
shown in figure~\ref{s}. Young stars are the main contributors to
the UV so since there is no star formation the elliptical galaxies
are weak in the UV. The amount of reprocessing in the galaxy
(figure~\ref{f}) is always small, at all times less than 5\% and
follows the mass of dust present in the galaxy (figure~\ref{c}b) so
it peaks at $\sim$ 8 Gyrs and then declines due to destruction from
the x-rays emitted by the hot gas.

Figure~\ref{r} shows that of the two simplifications the largest
effect on the SEDs, comes from assuming the dependence of the
dust-to-gas ratio on the metallicity will lead to large disparities
in the SED. This is because this assumption leads to large
differences in the dust to gas ratio with the results obtained from
the full CPM08 model (see figure~\ref{cc}b. As figure~\ref{r} shows
these disparities will have the largest impact in the FIR part of
the spectrum.
\section{Implications of this Work}
In this work only one likely chemical and dust evolution model, with
one star-formation history, has been followed for each morphological
type of galaxy based on average properties of the galaxy type as a
whole. It is therefore not possible to prove conclusively the
accuracy of the models, since it is unlikely that they will match
exactly the SED of any one specific galaxy. However, since the
chemical-evolution models are designed to match average properties,
the generated SEDs would be expected to match the general appearance
of many galaxies of the specific type and this is indeed what has
been found. In this paper plots have been presented which show
acceptable fits to local spirals, irregulars and ellipticals
simultaneously in their starbursting, post star-bursting and passive
phase. It should also be noted that although both models contain
many parameters, very little fine tuning has been performed to fit
observations. Instead the parameters have been set either to
physically reasonable values based on observations or to values
which, when used previously, have been found to give good fits to
observations. As a result, the fits are remarkably good and,
although far from conclusive, seem to validate both the chemical
evolution model used (the CPM08 model) and the stellar population
model adopted (GRASIL). Only two parameters were tuned in this work
to match the observations. For the irregular galaxies in order to
fit the MIR part of the SEDs of local dwarf irregulars the optical
depth of the molecular clouds (MCs) in which the stars are born,
needed to be reduced to a lower optical depth than the MCs in other
starforming galaxies. It is not possible to say whether the improved
fit requires a lower MC mass or a larger MC radius because the
optical depth is dependent on both the mass and the radius and is
given by: $M_{mc}/R_{mc}^{2}$. For the elliptical galaxies, to match
the large amount of reprocessing observed in the SEDs of the SCUBA
galaxies in the SHADES dataset, the time spent initially by the
young stars in the dense MCs was increased to a larger value than
that found for local normal starforming galaxies.

In the CPM08 paper it was found that the dust depletion pattern is
determined by the balance between destruction and accretion and
since neither process can be constrained accurately from the solar
neighborhood nothing definite can be learned about the rate of
either. However, in both the elliptical and irregular galaxies the
accretion rates should be negligible, hence, by studying such
environments it should in theory be possible to derive constraints
for the destruction rates of the different elements. The reasonable
fits to local galaxies support the values chosen.

The use of theoretical template SEDs in comparisons with photometry
has proved to be a useful tool in order to gain insights into high
redshift galaxies, in particular for determining a photometric
redshift (e.g. Rowan-Robinson et al. 2008). Among the templates
commonly used are those of well studied starburst galaxies in the
local universe, such as ARP 220 and M82. However there is evidence
that starburst galaxies at high redshift display different
characteristics to their low redshift counterparts. For example
showing higher specific star formation rates (star formation rate /
stellar mass) (Bauer et al. 2005) and have higher dust obscuration
rates (Hammer et al. 2005). It is therefore important to develop a
set of multi wavelength templates specific to the type of galaxy
expected at high redshift, particularly with the approaching launch
of the European Space Agency's (ESA) Herschel Space Observatory
(Pilbratt 2005), which will increase the wavelength coverage for
these objects. Our models deal with both the chemical and dust
evolution in these early galaxies in a self consistent way and in
this paper the SEDs generated have been shown to match high redshift
galaxies simultaneously in both the starbursting and post
starbursting stage. They should therefore provide ideal templates
for the analysis of these high redshift galaxies. Further work will
be carried out to generate and test a range of high redshift
templates.

Also investigated in this work are the potential errors you could
expect by introducing two common simplifications into the modeling
of galaxies: that the dust-to-gas mass ratio is proportional to the
metallicity, and that the dust composition in all galaxies is equal
to that derived for the Milky Way. These \emph{errors} are given as
the difference between our fiducial models with the detailed dust
treatment and the models using the two simplifications. They will
only be correct for the models presented here and not for any real
galaxy in particular. However the chemical evolution models have
been thoroughly tested against observations in previous works and in
this paper the SEDs have been shown to be broadly consistent with
the SEDs of the morphological type modeled. Therefore the
disparities presented in this work can be viewed as an example of
the magnitude of errors that could be expected if the
simplifications are adopted in spiral, irregular and elliptical
galaxies. In particular it has been shown that for spiral galaxies
the errors that could be expected are small and decrease with the
age of the galaxy, this is unsurprising since these simplifications
have been developed based on the Milky Way in the case of the
chemical composition and observations on local starforming galaxies
in the case of the dust-to-gas ratio. For irregular and elliptical
galaxies however it has been shown that potentially the disparities
introduced by adopting the two simplifications could be much larger.
Of the two simplifications it has been shown that in both galactic
types that the simplification likely to lead to the largest
disparities is assuming the dependency on the metallicity of the
dust-to-gas ratio and that the largest disparities would be expected
in the IR part of the spectrum.

The main limitation of this work is that only the total mass in each
of the main dust species is followed. The model does not calculate
in what proportions they combine with to make up dust grains and
what the size distribution of these dust species could be. As a
result a major assumption had to be introduced for the composition
of the dust grains, that the size distribution does not evolve and
all times will be identical to that calculated to match properties
of the Milky Way (S98 size distribution). This assumption was
relaxed in irregular galaxies where a dust composition consistent
with the SMC extinction curve was calculated and in young elliptical
galaxies where a dust composition consistent with the extinction
curve of a high redshift QSO was found. It was shown that adopting
the dust composition based on the SMC had only a small effect on the
SEDs. The effect of adopting the dust composition consistent with
the QSO in young elliptical galaxies had a large effect particularly
in the NIR and MIR. With the current observational data it was not
possible to put strong constraints on which size distribution is the
more realistic but the huge difference in the SEDs generated using
the two different distributions highlights how important using the
correct size distribution could be and how it could have large
consequences for surveys or any observations taken shortward of the
peak of the dust emission. Precisely the wavelength range which the
Herschel Space Observatory is set to operate in.
\section{Summary and Conclusions}
Starting from the models proposed in the CPM08 paper the chemical
and dust evolution models have been combined with the GRASIL
spectrophotometric population synthesis code in order to generate
SEDs for the solar neighborhood and irregular and elliptical
galaxies from the birth of the galaxy to an age of 13 Gyr.

The SEDs have been compared where possible to observed SEDs. A good
fit has been obtained for the spiral model when compared to local
spiral galaxies, the elliptical model has simultaneously fitted
observations of high-redshift starbursting (SCUBA) galaxies,
high-redshift post starburst galaxies and  local galaxies, while the
irregular galaxy model has been shown to match adequately the SEDs
of local irregular galaxies of magellanic type. The results are
particularly remarkable since very few of the GRASIL parameters were
fine tuned in order to obtain these results, instead they were
merely set to a combination of observationally motivated values or
to values which have been found to give good fits to observations in
previous works. The good fits give further indications that the
galaxy formation scenarios chosen for each morphological type in
order to fit the chemical properties of these galaxies are indeed
valid. In addition the good fits particularly to the local
elliptical and irregular galaxies suggest that the dust treatment
adopted in the CPM08 paper are reliable and in particular support
the destruction and accretion rates chosen. From these fits to
observations it has been shown that the optical depth in MCs in
irregular galaxies is likely to be lower than that of normal
starforming galaxies. It has also been found that in order to match
the observations of SCUBA galaxies the stars in the young extreme
starbursting galaxies initially spend a large period of time in
dense MCs before escaping.

It has been shown that SEDs generated using either the assumption
that the dust grain distribution in all environments is equal to
that of the Milky Way or the assumption that the dust-to-gas mass
ratio scales with the metallicity differ from those calculated using
a full dust evolution model. Of these assumptions it has been found
that assuming a dust-to-gas ratio will introduce the largest
disparities while the disparities introduced from an assumed
chemical composition are small, only having a potentially large
effect in the MIR, due to a change in the abundance of PAH
molecules. For spiral galaxies the disparities incurred by adopting
the assumptions are small and will decrease with age and therefore
increase with redshift, to a large part due to the normalisation of
the assumptions in this model to the results calculated for the
spiral galaxy at 13 Gyrs. The disparities introduced by adopting the
assumptions are more significant in the galaxies of the two other
morphologies at all ages particularly in elliptical galaxies and
therefore dust assumptions tuned to match those of the Milky Way
should be used with caution in these galaxies.

It has also been shown that while using a Milky Way size
distribution to model irregular galaxies should be a reasonable
simplification, for elliptical galaxies at high redshift such a
simplification could potentially lead to large errors.

The reasonable fit of the young elliptical galaxy to SCUBA galaxies
at a redshift of $\sim$ 2.2 and to a massive post starburst
elliptical at a redshift of $\sim$ 5 suggests that the starbursting
scenario adopted in this work is a reasonable one and that models
generated by these models could act as useful templates to fit to
high redshift galaxies.

This study can be thought of as a first step toward a more careful
comparison between galaxy formation scenarios and observed data
particularly in the spectral regions affected by dust emission. The
robustness of such a comparison has clearly been shown to be
affected by a proper treatment of dust evolution and reprocessing,
in a wavelength dependent way. A more rigorous dust treatment which
can be applied to galaxy formation models will be the subject of
future papers.

\section*{Acknowledgments}

We are indebted to an anonymous referee, whose insightful comments
have improved the quality of this paper. This work was supported
through a Marie Curie studentship for the 6th Framework Research and
Training network MAGPOP, contract number MRTN-CT-2004-503929.
%
%

%%%%%%%%%%%%%%%%%%%%%%%%%%%%%%%%%%%%%%%%%%%%%%%%%%%%%%%%%%%%%%%%%%%%%%%%%%%%%%%%%%%%%%%%%%%%%%%%%%%%%%%%%%%%%%%%%%%%%%

\bsp

\label{lastpage}


\begin{thebibliography}{10}
%
\bibitem[\protect\citeauthoryear{Bauer et al.} {2005}]{b1} Bauer, A. E., Drory, N., Hill, G. J.,  Feulner, G. 2005, ApJ, 621, L89
\bibitem[\protect\citeauthoryear{Baugh et al.} {2005}]{b1} Baugh C. M., Lacey C. G., Frenk C. S., Granato G. L., Silva
L., Bressan A., Benson A. J., Cole S., 2005, MNRAS, 356, 1191
\bibitem[\protect\citeauthoryear{Beelen et al.} {2006}]{b1} Beelen, A., Cox, P., Benford, D.J., Dowell, C.D., Kovacs, A., Bertoldi, F., Omont, A., Carilli,
C.L. 2006 ApJ, 642, 694
\bibitem[\protect\citeauthoryear{Bell et al.} {2005}]{b1} Bell et al. 2005 ApJ 625 23
\bibitem[\protect\citeauthoryear{Bertoldi et al.} {2003}]{b1} Bertoldi, F. et
al. 2003, 406, 55
\bibitem[\protect\citeauthoryear{Bianchi} {2007}]{b2} Bianchi, S. 2007, A\&A, 471, 765
\bibitem[\protect\citeauthoryear{Bianchi \& Schneider} {2007}]{b2} Bianchi, S., \& Schneider, R. 2007, MNRAS, 378, 973
\bibitem[\protect\citeauthoryear{Bianchi et al.} {2000}]{b2} Bianchi, S., Davies, J.I., Alton, P.B. 2000, A\&A, 359, 65
\bibitem[\protect\citeauthoryear{Bower et al.} {2006}]{b2} Bower R. G., Benson A. J., Malbon R., Helly J. C., Frenk C. S., Baugh
C. M., Cole S., Lacey C. G., 2006, MNRAS, 370, 645
\bibitem[\protect\citeauthoryear{Bradamamente et al. } {1998}]{b2} Bradamante F., Matteucci F., D'Ercole A., 1998, A\&A, 337, 338
\bibitem[\protect\citeauthoryear{Brinchmann \& Ellis} {2000}]{b2}Brinchmann, J., \& Ellis, R. S. 2000, ApJ, 536, L77
\bibitem[\protect\citeauthoryear{Bressan et al.} {1994}] {b2} Bressan, A., Chiosi, C., Fagotto, F. 1994, ApJS, 94, 63
\bibitem[\protect\citeauthoryear{Bressan et al. } {1998}]{b2} Bressan A., Granato G. L., Silva L., 1998, A\&A, 332, 135
\bibitem[\protect\citeauthoryear{Calura, Pipino \& Matteucci}{2008}]{b1} Calura, F., Pipino, A., Matteucci, F. 2008, A\&A, 479, 669
\bibitem[\protect\citeauthoryear{Chiappini et. al } {1997}]{b3} Chiappini C., Matteucci F., Gratton R., 1997, ApJ, 477,
765
\bibitem[\protect\citeauthoryear{Chiappini et. al } {2001}]{b4} Chiappini C., Matteucci F., Romano D., 2001, ApJ, 554,
1044
\bibitem[\protect\citeauthoryear{Clayton et al.} {2001}]{b5} Clayton, Geoffrey C., Green, J., Wolff, Michael J., Zellner, Nicolle E. B.,
Code, A. D., Davidsen, Arthur F., WUPPE Science Team, HUT Science
Team, 1996, ApJ, 460 313
\bibitem[\protect\citeauthoryear{Clements et al.} {2008}]{b5}
Clements et al., 2008, MNRAS, 387, 247
\bibitem[\protect\citeauthoryear{Cowie et al.} {1996}]{b5} Cowie, L. L., Songaila, A., Hu, E. M., \& Cohen, J. G. 1996, AJ, 112, 839
\bibitem[\protect\citeauthoryear{Dale \& Helou} {2002}]{b5}Dale, D. A., Helou, G. 2002, ApJ, 576, 159
\bibitem[\protect\citeauthoryear{Dale et al.} {2001}]{b5} Dale, D.A., Helou, G., Contursi, A., Silbermann, N.A., Kolhatkar, S. 2001,  ApJ, 549, 215
\bibitem[\protect\citeauthoryear{Dale et al.} {2007}]{b5} Dale D.A., Gil de Paz A., Gordon
K.D., et al. 2007 ApJ, 655, 863
\bibitem[\protect\citeauthoryear{de Vaucouleurs et al.} {1991}]{b6} de Vaucouleurs G., de Vaucouleurs A., Corwin H. G., Jr, Buta R. J., Paturel
G., Fouque P., 1991, Third Reference Catalogue of Bright Galaxies,
Vols 1–3. Springer-Verlag, Berlin
\bibitem[\protect\citeauthoryear{Desert et al.} {1990}]{b6} Desert, F.-X, Boulanger, F., Puget, J.L. 1990, A\&A, 237, 215
\bibitem[\protect\citeauthoryear{Di Matteo et al.} {2005}]{b6} Di Matteo, T., Springel, V., Hernquist, L.
2005, Nature, 433, 604
\bibitem[\protect\citeauthoryear{Dopita} {2005}]{b6} Dopita, M.A., 2005, AIPC, 761, 203D
\bibitem[\protect\citeauthoryear{Draine \& Lee} {1984}]{b6} Draine B.T., Lee., H.M., 1984, ApJ, 285, 89
\bibitem[\protect\citeauthoryear{Draine \& Lee} {2001}]{b7} Draine B.T., Li., A., 2001, ApJ, 551, 807
\bibitem[\protect\citeauthoryear{Draine \& Lee} {2007}]{b8} Draine B.T., Li., A., 2007, ApJ, 657, 810
\bibitem[\protect\citeauthoryear{Draine et al} {2007}]{b8} Draine B. T. et al., 2007, ApJ, 663, 866
\bibitem[\protect\citeauthoryear{Dunne et al} {2003}]{b8} Dunne L., Eales S., Ivison R., Morgan H., Edmunds M., 2003,
Nat, 424, 285
\bibitem[\protect\citeauthoryear{Dunne et al} {2008}]{b8} Dunne, L., Maddox, S. J., Ivison, R. J., Rudnick, L., DeLaney,
T. A., Matthews, B. C., Gomez, H. L., Eales, S. A., Crowe, C. M.,
Dye, S. arXiv0809.0887D, 2008
\bibitem[\protect\citeauthoryear{Dwek}{1998}]{b9} Dwek, E. 1998, ApJ, 501, 643
\bibitem[\protect\citeauthoryear{Dwek et al}{2007}]{b9} Dwek, E. Galliano, F., Jones, A.P. 2007, ApJ, 662, 927
\bibitem[\protect\citeauthoryear{Eales et al}{2000}]{b9} Eales S., Lilly S., Webb T., Dunne L., Gear W., Clements
D., Yun M., 2000, AJ, 120, 2244
\bibitem[\protect\citeauthoryear{Efstathiou \$ Rowan-Robinson, Michael }{2003}]{b9} Efstathiou A., Rowan-Robinson M., 2003, MNRAS, 343, 322
\bibitem[\protect\citeauthoryear{Efstathiou, Rowan-Robinson M \& Siebenmorgan}{2000}]{b9} Efstathiou A., Rowan-Robinson M., Siebenmorgan R., 2000, MNRAS, 313,
734
\bibitem[\protect\citeauthoryear{Engelbracht et al}{2005}]{b9} Engelbracht, C. W., et al. 2005, ApJ, 628, L29
\bibitem[\protect\citeauthoryear{Fran\c{c}ois et al}{2004}]{b9} Fran\c{c}ois, P., Matteucci F., Cayrel R., Spite M., Spite F.,
Chiappini C., 2004, A\&A, 421, 613
\bibitem[\protect\citeauthoryear{Galliano et al}{2003}]{b9} Galliano, F., Madden, S. C., Jones, A. P., Wilson, C. D., Bernard, J.-P., Le Peintre, F. 2003,
A\&A, 407, 159
\bibitem[\protect\citeauthoryear{Galliano et al}{2005}]{b9} Galliano, F., Madden, S. C., Jones, A. P., Wilson, C. D., Bernard, J.-P. 2005, A\&A, 434, 867
\bibitem[\protect\citeauthoryear{Galliano et al}{2008}]{b9} Galliano, F., Dwek, E., Chanial,
P. 2008, ApJ, 672, 214
\bibitem[\protect\citeauthoryear{Granato et al}{2000}]{b9} Granato G. L. et al., 2000, ApJ, 542, 710
\bibitem[\protect\citeauthoryear{Granato et al}{2004}]{b9} Granato G. L., De Zotti G., Silva L., Bressan A., Danese L., 2004, ApJ, 600,
580
\bibitem[\protect\citeauthoryear{Guzman et al}{1997}]{b9} Guzman, R., Gallego, J., Koo, D. C., Phillips, A. C., Lowenthal, J. D., Faber,
S. M., Illingworth, G. D., \& Vogt, N. P. 1997, ApJ, 489, 559
\bibitem[\protect\citeauthoryear{Hammer et al}{2005}]{b9} Hammer, F., Flores, H., Elbaz, D., Zheng, X. Z., Liang, Y. C., Cesarsky,
C. 2005, A\&A, 430, 115
\bibitem[\protect\citeauthoryear{Hauser \& Dwek}{2001}]{b9} Hauser, M. G., Dwek, E. 2001, ARA\&A, 39, 249
\bibitem[\protect\citeauthoryear{Hopkins \& Hernquist}{2008}]{b9} Hopkins F. P., \&
Hernquist L., 2008 arXiv:0809.3789v1
\bibitem[\protect\citeauthoryear{Hunt et al.}{2005}]{b9} Hunt, L., Bianchi, S., Maiolino, R. 2005, A\&A, 434, 849
\bibitem[\protect\citeauthoryear{Itoh H}{1989}]{b9} Itoh H., 1989, PASJ, 41, 853
\bibitem[\protect\citeauthoryear{Juneau, S et al}{2005}]{b9} Juneau, S., et al. 2005, ApJ, 619, L135
\bibitem[\protect\citeauthoryear{Kaviraj et al.}{2007}]{b9} Kaviraj S. et al., 2007, ApJS, 173, 619
\bibitem[\protect\citeauthoryear{Kennicutt}{1989}]{b9} Kennicutt, R.C., 1989, ApJ, 344, 685
\bibitem[\protect\citeauthoryear{Kennicutt}{1998}]{b9} Kennicutt, R. C. 1998, ApJ, 498, 541
\bibitem[\protect\citeauthoryear{Kimura, Mann \& Jessberger}{2003}]{b9} Kimura, H., Mann, I., \& Jessberger, E. K. 2003, ApJ, 582, 846
\bibitem[\protect\citeauthoryear{Kodama et al.}{2004}]{b9} Kodama T., Yamada T., Akiyama M. et al., 2004, MNRAS, 350, 1005
\bibitem[\protect\citeauthoryear{Krause et al.}{2004}]{b9} Krause O., Birkmann S. M., Rieke G. H., Lemke D., Klaas U.,
Hines D. C., Gordon K. D., 2004, Nat, 432, 596
\bibitem[\protect\citeauthoryear{Laor \& Draine et al}{2005}]{b9} Laor, A., Draine, B. T. 1993, ApJ, 402, 441
\bibitem[\protect\citeauthoryear{Lilly}{1999}]{b9} Lilly, S. J. et al. 1999, ApJ, 518, 641
\bibitem[\protect\citeauthoryear{Lisenfeld \& Ferrara}{1998}]{b10} Lisenfeld U., Ferrara A., 1998, ApJ, 496, 145
\bibitem[\protect\citeauthoryear{Madden}{2005}]{b10} Madden S.C., 2005, AIPC, 761, 223M
\bibitem[\protect\citeauthoryear{Madden et. al}{2006}]{b10} Madden, S. C., Galliano, F., Jones, A. P., \& Sauvage, M. 2006, A\&A, 446, 877
\bibitem[\protect\citeauthoryear{Maiolino} {2004}]{b11} Maiolino,
R., Schneider, R., Oliva, E., Bianchi S., Ferrara, A., Mannucci, F.,
Pedani, M., Roca Sogorb, M. 2004 Nature, 431, 553
\bibitem[\protect\citeauthoryear{Maiolino} {2006}]{b11} Maiolino, R., et al. 2006, Mem. Soc. Astron. Italiana, 77, 643
\bibitem[\protect\citeauthoryear{Marigo et al.} {2008}]{b11} Marigo P., Girardi L., Bressan A., Groenewegen M. A. T., Silva L., Granato
G. L., 2008, A\&A, 482, 883
\bibitem[\protect\citeauthoryear{Marconi et al.} {2004}]{b11} Marconi, A., Risaliti, G., Gilli, R., Hunt, L. K., Maiolino, R., \& Salvati, M.
2004, MNRAS, 351, 169
\bibitem[\protect\citeauthoryear{Mathis, Rumple \& Nordsieck} {1977}]{b11} Mathis, J.S. Rumple, W. Nordsieck K.H., 1977, ApJ, 217, 103
\bibitem[\protect\citeauthoryear{Matteucci} {1994}]{b12} Matteucci F., 1994, A\&A, 288, 57
\bibitem[\protect\citeauthoryear{Matteucci \& Tornamb\`{e}} {1987}]{b13} Matteucci F., Tornamb\`{e} A., 1987,
A\&A, 185, 51
\bibitem[\protect\citeauthoryear{McKee et al} {1989}]{b13} McKee C. F., 1989, in Allamandola L. J., Tielens A. G. G. M., eds, Interstellar
Dust, Proc. IAU Symposium 135. Kluwer, Dordrecht, p. 431
\bibitem[\protect\citeauthoryear{Mclure \&
Dunlop} {2002}]{b13} McLure R. J., Dunlop J. S., 2002, MNRAS, 331,
795
\bibitem[\protect\citeauthoryear{Misiriotis et al} {1996}]{b13} Misiriotis, A.; Xilouris, E. M.; Papamastorakis, J.; Boumis, P.; Goudis, C.
D. 2006, A\&A, 459,113
\bibitem[\protect\citeauthoryear{Mobasher}{2005}]{b9} Mobasher, B. et al. 2005, ApJ, 635, 832
\bibitem[\protect\citeauthoryear{Morgan et al} {2003}]{b13} Morgan
H.L., Edmunds M.G., 2003, MNRAS, 343, 427
\bibitem[\protect\citeauthoryear{Nagashima et al.}{2005}]{b9} Nagashima, M., Lacey, C.
G., Okamoto, T., et al. 2005, MNRAS, 363L, 31
\bibitem[\protect\citeauthoryear{Noeske et al} {2007}]{b13} Noeske K. G., et al., 2007, ApJL, 660, L43
\bibitem[\protect\citeauthoryear{Nozawa et al} {2007}]{b13} Nozawa, T., Kozasa, T., Habe, A., Dwek, E., Umeda, H., Tominaga, N., Maeda,
K., \& Nomoto, K. 2007, ApJ, 666, 955
\bibitem[\protect\citeauthoryear{O'Halloran et. al}
{2006}]{b14} O'Halloran, B., Satyapal, S., \& Dudik, R. P. 2006,
ApJ, 641, 795
\bibitem[\protect\citeauthoryear{Panuzzo et
al.} {2002}]{b14} Panuzzo, P. et al. 2007, ApJ, 656, 206
\bibitem[\protect\citeauthoryear{Pilbratt} {2005}]{b14} Pilbratt G. L., 2005, in Wilson A., ed., ESA SP-577, Proc. Dusty and
Molecular Universe. ESA, Noordwijk, p. 3
\bibitem[\protect\citeauthoryear{Pipino et al.} {2002}]{b14} Pipino A., Matteucci F., Borgani S., Biviano A., 2002, NewA, 7, 227
\bibitem[\protect\citeauthoryear{Pipino et al.} {2005}]{b15} Pipino A., Kawata D., Gibson B.k., Matteucci F.,  2005, A\&A, 434, 553
\bibitem[\protect\citeauthoryear{Piovan et al.} {2006}]{b15} Piovan, L., Tantalo, R., Chiosi,
C., 2006, MNRAS, 370, 1454
\bibitem[\protect\citeauthoryear{Popescu \& Tuffs} {2005}]{b1} Popescu, C.C., Tuffs, R.J., 2005 AIPC, 761, 155P
\bibitem[\protect\citeauthoryear{Renzini} {2006}]{b15} Renzini A., 2006, ARA\&A, 44, 141
\bibitem[\protect\citeauthoryear{Robson et al.} {2004}]{b15} Robson,
I., Priddey, R.S., Isaak, K.G., McMahon, R.G. 2004, MNRAS, 351, L29
\bibitem[\protect\citeauthoryear{Scalo} {1986}]{b9} Scalo J. M., 1986, FCPh, 11, 1
\bibitem[\protect\citeauthoryear{Shankar et al.} {2005}]{b9}  Shankar, F., Salucci, P., Granato, G. L., de Zotti, G., \& Danese, L. 2005, in
Growing Black Holes: Accretion in a Cosmological Context, ed. A.
Merloni, S. Nayakshin, \& R. A. Sunyaev ( Berlin: Springer), 470
\bibitem[\protect\citeauthoryear{Silva et al.}{1998}]{b10} Silva L., Granato GL., Bressan A., Danese L., 1998, ApJ, 509, 103
\bibitem[\protect\citeauthoryear{Soifer \& Neugebauer} {1991}]{b9} Soifer B.T., Neugebauer G. 1991. Astron. J.
101:354–61
\bibitem[\protect\citeauthoryear{Somerville} {2004}]{b9} Somerville, R. S. 2004, in Multiwavelength Mapping of Galaxy Formation and
Evolution, ed. R. Bender \& A. Renzini ( Berlin: Springer), 131
\bibitem[\protect\citeauthoryear{Sugerman et al.} {2006}]{b9}
Sugerman et al. 2006, Sci, 313, 196
\bibitem[\protect\citeauthoryear{Swinbank et al.} {2008}]{b9}
Swinbank et al. 2008, MNRAS in press, (arXiv:0809.0973)
\bibitem[\protect\citeauthoryear{Temi et al.} {2007}]{b9} Temi P., Brighenti F., Mathews W., 2007, ApJ, 666, 222
\bibitem[\protect\citeauthoryear{Thomas et al.} {2005}]{b9} Thomas, D., Maraston, C., Bender, R., Mendes de Oliveira, C. 2005, ApJ, 621, 673
\bibitem[\protect\citeauthoryear{Tuffs et al.} {2004}]{b9} Tuffs R. J., Popescu C. C., V¨olk H. J., Kylafis N. D., Dopita
M. A., 2004, A\&A, 419, 821
\bibitem[\protect\citeauthoryear{van den Hoek et al.} {1997}]{b9} van den Hoek L. B.  Groenwegen M. A. T., 1997, A\&AS, 123, 305
\bibitem[\protect\citeauthoryear{Vega et al.} {2004}]{b9} Vega, O., Silva, L., Panuzzo, P., et al. 2005, MNRAS, 364, 1286
Evolution, ed. J. D. Lowenthal, D. H. Hughes (Singapore:World
Scientific Publishing), 103
\bibitem[\protect\citeauthoryear{Walter et. al } {2007}]{b19} Walter, F., et al. 2007, ApJ, 661, 102
\bibitem[\protect\citeauthoryear{Weingartner \& Draine} {2001}]{b9} Weingartner, J. C., Draine, B. T. 2001, ApJ, 548, 296
\bibitem[\protect\citeauthoryear{Wiklind et al.} {2008}]{b9} Wiklind, T., et al. 2008, ApJ, 676, 781
\bibitem[\protect\citeauthoryear{Wilson et al.} {2008}]{b9} Wilson T. L., Batrla W., 2005, A\&A, 430, 561
\bibitem[\protect\citeauthoryear{Wu et al.} {2006}]{b9} Wu, Y., Charmandaris, V., Hao, L., Brandl, B. R., Bernard-Salas,
J., Spoon, H. W. W., \& Houck, J. R. 2006, ApJ, 639, 157
\bibitem[\protect\citeauthoryear{Xilouris et al.} {1997}]{b15} Xilouris, E.M., Kylafis, N.D., Papamastorakis, J., Paleologou, E.V., Haerendel, G 1997, A\&A, 325, 135
\bibitem[\protect\citeauthoryear{Xilouris et al.} {1998}]{b15} Xilouris, E.M. et al., 1998, A\&A, 344, 868
\bibitem[\protect\citeauthoryear{Xilouris et al.} {1999}]{b15} Xilouris, E.M., Kylafis, N.D., Paleologou, E.V., Papamastorakis, J. 1999, A\&A, 344, 868
\bibitem[\protect\citeauthoryear{Zhukovska, Gail \& Trieloff} {2007}]{b16} Zhukovska S., Gail H.P., Trieloff M., 2004, A\&A in press
\bibitem[\protect\citeauthoryear{Zubko et al.} {2004}]{b17} Zubko, V., Dwek, E., Arendt R.G., 2004, ApJS, 152, 211
%
\end{thebibliography}
\end{document}